\documentclass[aps,prx,twocolumn,nopacs,superscriptaddress]{revtex4-2}
\usepackage[utf8]{inputenc}
\usepackage[T1]{fontenc}

\usepackage{graphicx}  % needed for figures
\usepackage{dcolumn}   % needed for some tables
\usepackage{bm}        % for math
\usepackage{amssymb}   % for math
\usepackage{amsmath}
\usepackage{units}
\usepackage[english]{babel}
\usepackage[dvipsnames]{xcolor}
\usepackage{xfrac}
\usepackage{multirow}
\usepackage{stackrel}

\usepackage[%
colorlinks=true,
urlcolor=RoyalBlue,
linkcolor=RoyalBlue,
citecolor=RoyalBlue,
]{hyperref}
\usepackage{orcidlink}

\usepackage{vmargin}
\setpapersize{A4} \oddsidemargin20mm \textwidth170mm \topmargin10mm
\usepackage{sidecap}
\sidecaptionvpos{figure}{t}  

\usepackage[type1]{libertine}                                        % Linux Libertine für zweispaltige Texte
\usepackage{textcomp}% Required to get special symbols
\usepackage[scaled=.85]{beramono}% Typewriter font
\usepackage[libertine,cmintegrals,cmbraces,vvarbb,slantedGreek]{newtxmath}
\usepackage[scr=boondoxo]{mathalfa}% Extra math symbols
\usepackage{bm}% Extra bold faces
\usepackage[lf]{carlito}

\usepackage{braket}
\usepackage{bm}

\makeatletter
\DeclareRobustCommand{\cev}[1]{%
  \mathpalette\do@cev{#1}%
}
\newcommand{\do@cev}[2]{%
  \fix@cev{#1}{+}%
  \reflectbox{$\m@th#1\vec{\reflectbox{$\fix@cev{#1}{-}\m@th#1#2\fix@cev{#1}{+}$}}$}%
  \fix@cev{#1}{-}%
}
\newcommand{\fix@cev}[2]{%
  \ifx#1\displaystyle
    \mkern#21mu
  \else
    \ifx#1\textstyle
      \mkern#21mu
    \else
      \ifx#1\scriptstyle
        \mkern#20mu
      \else
        \mkern#20mu
      \fi
    \fi
  \fi
}
\makeatother

\newcommand{\spinhalf}{spin-\sfrac{1}{2}}

\renewcommand{\Re}{\text{Re}}

\newcommand{\panel}[1]{(#1)}
\newcommand{\panelcaption}[1]{(#1)}
\newcommand{\panelsubcaption}[1]{(#1)}

\begin{document}

\title{Theory of Electron Spin Resonance in Scanning Tunneling Microscopy}

\author{Christian R. Ast\,\orcidlink{0000-0002-7469-1188}}
\email[Corresponding author; electronic address:\ ]{c.ast@fkf.mpg.de}
\affiliation{Max-Planck-Institut f\"ur Festk\"orperforschung, Heisenbergstraße 1, 70569 Stuttgart, Germany}
\author{Piotr Kot}
\affiliation{Max-Planck-Institut f\"ur Festk\"orperforschung, Heisenbergstraße 1, 70569 Stuttgart, Germany}
\author{Maneesha Ismail}
\affiliation{Max-Planck-Institut f\"ur Festk\"orperforschung, Heisenbergstraße 1, 70569 Stuttgart, Germany}
\author{Sebastián de-la-Pe\~na}
\affiliation{Departamento de F\'{\i}sica Te\'orica de la Materia Condensada and
Condensed Matter Physics Center (IFIMAC), Universidad Aut\'onoma de Madrid, 28049 Madrid, Spain}
\author{Antonio I. Fern\'andez-Dom\'{\i}nguez\,\orcidlink{0000-0002-8082-395X}}
\affiliation{Departamento de F\'{\i}sica Te\'orica de la Materia Condensada and
Condensed Matter Physics Center (IFIMAC), Universidad Aut\'onoma de Madrid, 28049 Madrid, Spain}
\author{Juan Carlos Cuevas\,\orcidlink{0000-0001-7421-0682}}
\affiliation{Departamento de F\'{\i}sica Te\'orica de la Materia Condensada and
Condensed Matter Physics Center (IFIMAC), Universidad Aut\'onoma de Madrid, 28049 Madrid, Spain}

\date{\today}

\begin{abstract}
Electron spin resonance (ESR) spectroscopy in scanning tunneling microscopy (STM) has enabled probing the electronic structure of single magnetic atoms and molecules on surfaces with unprecedented energy resolution, as well as demonstrating coherent manipulation of single spins. Despite this remarkable success, the field could still be greatly advanced by a more quantitative understanding of the ESR-STM physical mechanisms. Here, we present a theory of ESR-STM which quantitatively models not only the ESR signal itself, but also the full background tunneling current, from which the ESR signal is derived. Our theory is based on a combination of Green's function techniques to describe the electron tunneling and a quantum master equation for the dynamics of the spin system along with microwave radiation interacting with both the tunneling current and the spin system. We show that this theory is able to quantitatively reproduce the experimental results for a \spinhalf\ 
system (TiH molecules on MgO) across many orders of magnitude in tunneling current, providing access to the relaxation and decoherence rates that govern the spin dynamics due to intrinsic mechanisms and to the applied bias voltage. More importantly, our work establishes that: (i) sizable ESR signals, which are a measure of microwave-induced changes in the junction magnetoresistance, require surprisingly high tip spin polarizations, (ii) the coupling of the magnetization dynamics to the microwave field gives rise to the asymmetric ESR spectra often observed in this spectroscopy. Additionally, our theory provides very specific predictions for the dependence of the relaxation and decoherence times on the bias voltage and the tip-sample distance. Finally, with the help of electromagnetic simulations, we find that the transitions in our ESR-STM experiments, in which the tunnel junction is irradiated by a nearby microwave antenna, can be driven by the ac magnetic field at the junction.
\end{abstract}

\maketitle  

\section{Introduction}

\begin{SCfigure*}
    \centering
    \includegraphics[width=0.66\textwidth]{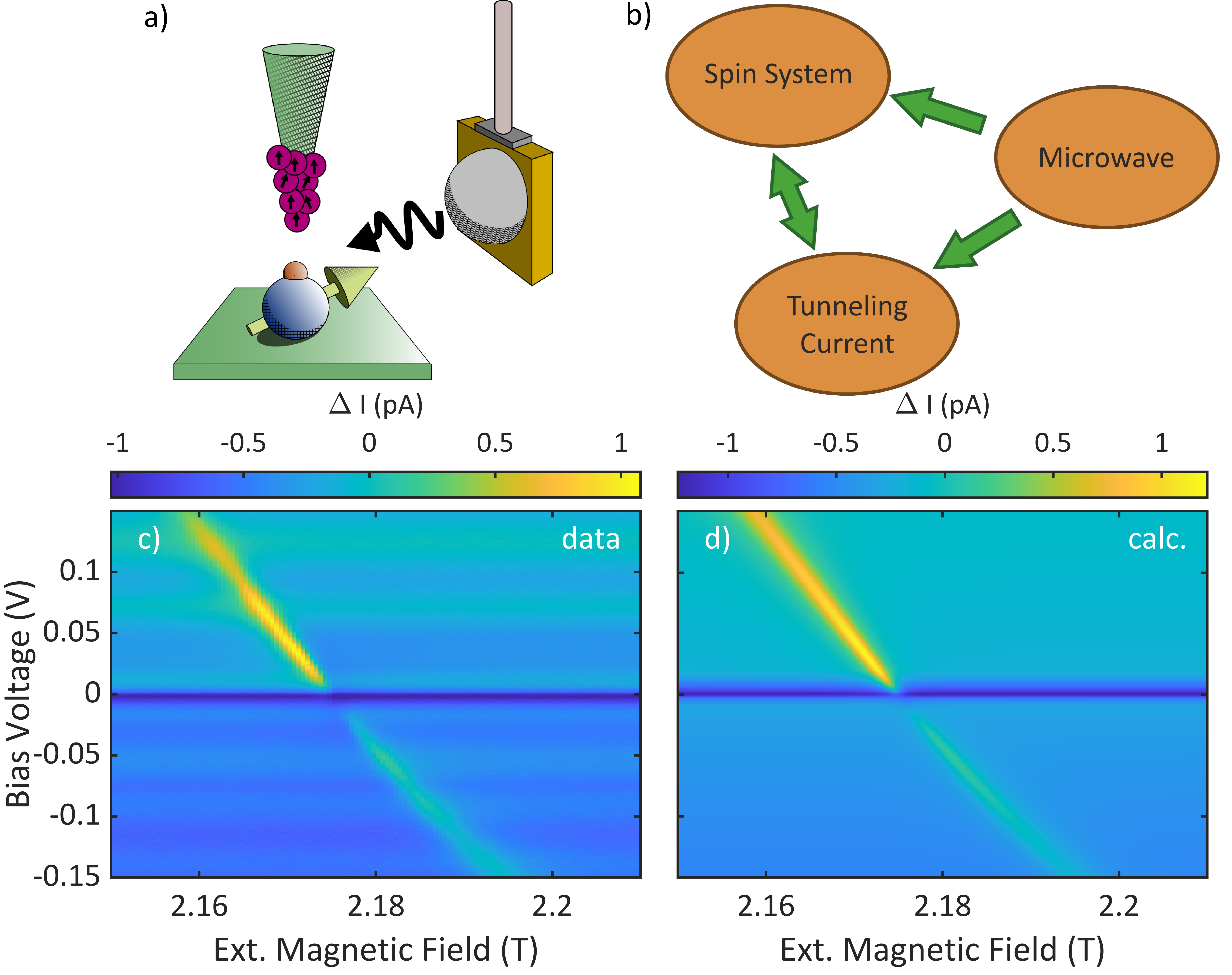}
    \centering
    \caption{\textbf{Modeling the ESR-STM data:} \panelcaption{a} Schematic drawing of the ESR tunnel junction. The microwave 
    is supplied from an antenna radiating into the vacuum towards the tunnel junction. \panelcaption{b} The tunneling current 
    and the spin system influence each other, while both interact with the microwave. The tunneling current in our theory is modeled 
    by nonequilibrium Green's functions and the spin system dynamics is described by a master equation to calculate the density matrix. 
    \panelcaption{c} Measured ESR signal as function of bias voltage and magnetic field at constant microwave frequency 
    (61\,GHz) in TiH molecules on MgO. \panelcaption{d} Calculated ESR signal using parameters fitted to the experimental data 
    in \panelsubcaption{c} and the theory presented in this work.}
    \label{fig:Fig1}
\end{SCfigure*}

Electron spin resonance (ESR), also referred to as electron paramagnetic resonance, is a widely used spectroscopy and 
imaging technique in chemistry, biology, and condensed matter physics to characterize systems with unpaired electrons 
\cite{abragam1970electron}. The main advantage of ESR is its high energy resolution, only limited by the decoherence time of 
magnetic excitations, while its spatial resolution is limited by the magnetic field gradients that can be implemented (resulting 
in a resolution of the order of 100 $\mu$m$^3$ \cite{Fratila2011}). In 2015, Baumann and coworkers reported the first 
convincing implementation of ESR in the context of scanning tunneling microscopy (STM) by addressing single magnetic atoms 
on a surface using an oscillating electric field (20-30 GHz) \cite{baumann_electron_2015}. In ESR-STM, the lifting of the spin 
degeneracy by means of an external magnetic field and the spin excitation through an external microwave source is done in analogy 
to conventional ESR. The detection of the spin state, however, is realized through the measurement of the tunneling current in 
an appropriately spin-polarized tip, which ensures atomic scale spatial resolution, and is particular to STM. This is 
schematically shown in Fig.\ \ref{fig:Fig1}\panel{a}. Since the first demonstration, many different groups have reported 
experimental ESR-STM studies \cite{natterer2017reading,choi2017atomic,willke_probing_2018,willke2018hyperfine,
bae_enhanced_2018,yang2018electrically,willke2019tuning,willke_magnetic_2019,yang2019tuning,yang_coherent_2019,seifert2020longitudinal,
van2021scanning,steinbrecher2021quantifying,veldman2021free,kim2022anisotropic,kovarik2022electron,zhang_electron_2022,kot_electric_2022,
phark2023double}. Thus, for instance, ESR-STM has been used to measure the hyperfine interaction in Ti atoms on a MgO surface
\cite{willke2018hyperfine}, to achieve the coherent spin manipulation of individual atoms on a surface \cite{yang_coherent_2019}, 
to do magnetic resonance imaging of single atoms on a surface \cite{willke_magnetic_2019}, to demonstrate the electric control of 
spin transitions at the atomic scale \cite{kot_electric_2022}, and to do basic quantum computing operations with single atoms 
on a surface \cite{phark2023double}.

\sidecaptionvpos{figure}{c}

However, the physical mechanisms underlying ESR-STM are still under debate, for example, what drives the transitions between the 
magnetic states in an all-electric ESR-STM experiment. Many different mechanisms have been suggested \cite{baumann_electron_2015,
seifert2020longitudinal,lado2017exchange,ferron2019single,galvez2019cotunneling,reina_all-electric_2021,reina2023many}, but there 
is no real consensus on this issue (for a detailed account of this debate, see Ref.~\onlinecite{Delgado_Lorente_2021}). Further 
points that need more clarification include, but are not limited to, what constitutes a suitable tip for acquiring ESR signals
\cite{willke_magnetic_2019,Seifert2021,Rodriguez2023}, how to extract relaxation and decoherence times of the spin excitations 
from continuous mode ESR spectra, how the bias voltage influences the ESR mechanis, and why some ESR signals are more asymmetric 
than others. Also, the role of the substrate is still poorly understood and most of the experiments have been done in the same 
type of substrate (MgO), although other substrates (NaCl) have been successfully explored recently \cite{kawaguchi_spatially_2023}. 
To address these questions, it would be highly desirable to have an ESR-STM theory able to quantitatively reproduce the measured 
signals. The goal of this work is to provide such a theory and answer many of these open questions.

Part of the problem in developing an ESR-STM theory is that the interaction of the tunneling current with the spin system does 
not just probe the spin state, but also has a strong back-action on the dynamics of the spin state itself \cite{loth_controlling_2010,delgado_spin-transfer_2010,delgado_spin_2010}. 
In ESR-STM, the microwaves introduce an additional component that interacts both with the spin system as well as the tunneling 
current. The complex interplay between the spin system, the tunneling current and the microwave is schematically shown in 
Fig.~\ref{fig:Fig1}\panel{b}. Thus, a quantitative understanding of the tunneling current has to take these three components 
into account along with their interactions. Here, we present a theory that quantitatively describes ESR-STM spectra. Our theory 
is based on a combination of nonequilibrium Green's function techniques for the calculation of the tunneling current and quantum 
master equations for the description of the spin dynamics of the magnetic system. This theory takes into account the role of microwaves 
both in the tunneling processes and in the spin dynamics. To test our theory we also present here experimental ESR-STM spectra 
measured in TiH molecules on a MgO substrate, a known \spinhalf\ system, as a function of the external magnetic field and the bias 
voltage, see Fig.~\ref{fig:Fig1}\panel{c}. Our theory is able to quantitatively reproduce all the salient features of the experimental 
results across many orders of magnitude in current, see Fig.~\ref{fig:Fig1}\panel{d} and discussion below. This quantitative 
agreement allows us in turn to unambiguously determine the relaxation and decoherence times of the system. Moreover, we supplement 
our theory with electromagnetic simulations to determine the magnitude of the electric and magnetic field in the region of the tunnel 
junction and to provide further insight as to what drives the ESR transitions in our experiments. Our study leads to a 
number of important conclusions for the field of ESR-STM, among which are the following ones: 
\begin{enumerate}
\item The transitions in our ESR-STM experiments are driven by the ac magnetic field that accompanies the microwave radiation.

\item We confirm that the measured ESR signal is predominantly due to the change in magnetoresistance resulting from 
 microwave-induced transitions. We provide an analytical formula for the ESR spectra that corroborates the phenomenological 
 expressions often employed in the literature. Our formula includes voltage and current dependence of the relaxation and 
 decoherence times. It also reveals the need for a high spin polarization of the tips to obtain sizeable ESR signals.
 
\item A misalignment of the spins between the magnetic impurity and the STM tip results in a coupling of the magnetization 
dynamics to the microwave field, known as homodyne detection, which gives rise to an asymmetry in the ESR spectra often observed 
in this spectroscopy. This mechanism also explains our experimental ESR signal near zero bias voltage.
\end{enumerate}

The rest of this manuscript is organized as follows. In Sec.~\ref{sec-general} we present our theory to describe the ESR-STM 
spectra in an arbitrary spin system. Then, in Sec.~\ref{sec-TLS} we apply this theory to the case of a \spinhalf\ system. 
Section~\ref{sec-exp} is devoted to a brief description of our experimental ESR-STM results obtained in TiH molecules on a MgO 
substrate and used to test our theory. We present in Sec.~\ref{sec-comp} an analysis of these experimental results in light of 
our theory and explain how the different rates and time scales can be extracted from such an analysis. In Sec.~\ref{sec-EM}, 
we present electromagnetic simulations that aim at determining the magnitude of the electromagnetic fields at the junction region 
with the goal to clarify what determines the transitions in our ESR-STM experiments. Finally, we summarize the main conclusions 
of this work in Sec.~\ref{sec-conclusions}. We also include two appendices to provide more details about our experiment, as well 
as to report some analytical results for the \spinhalf\ system.

\section{General ESR-STM theory} \label{sec-general}

In this section we present our general ESR-STM theory, which we will apply to a \spinhalf\ system in the next section. The goal 
of this ESR-STM theory is to describe the electrical current through a spin system (or magnetic impurity) deposited on a substrate 
and addressed by an STM tip in the presence of a microwave field, see Fig.~\ref{fig:Fig1}(a). Our approach is based on the use 
of a tunneling Hamiltonian that describes both the spin-flip processes that can occur in this system and the microwave-assisted 
processes induced by the microwave field. We use this tunneling Hamiltonian in combination with linear response theory to express 
the current in terms of spin correlation functions of the spin system. Those correlation functions are the determined with the help 
of quantum master equations that describe the dynamics of the spin system. In the following, we explain how these different 
ingredients are combined to provide an ESR-STM theory for an arbitrary spin system. 

\subsection{Tunneling Hamiltonian} \label{sec-HT}

We model the tunnel junction depicted in Fig.~\ref{fig:Fig1}(a) by a left and a right reservoir $H_\text{L,R}$, which correspond 
to the substrate and the STM tip, along with a tunnel coupling $H_\text{T}$, such that the total Hamiltonian becomes
\begin{equation}
    H = H_\text{L} + H_\text{R} + H_\text{T}.
\end{equation}
We do not specify the left and right reservoirs any further other than by the corresponding spin-dependent quasiparticle 
creation and annihilation operators $c^\dagger_{L,R\sigma}$ and $c_{L,R\sigma}$, respectively, where $\sigma$ denotes the 
spin index $\uparrow$ or $\downarrow$ for \spinhalf\ particles. We describe the effective tunnel coupling between the two 
electrodes as follows
\begin{equation} \label{eq-HT}
H_\text{T}(\tau) = t e^{i\phi(\tau)} \sum_{\sigma,\sigma^{\prime}} c^{\dagger}_{L\sigma}(\tau) \left( \delta_{\sigma \sigma^{\prime}} + 
\frac{\lambda}{\hbar} \boldsymbol{\sigma}_{\sigma \sigma^{\prime}} \cdot \boldsymbol{S}(\tau) \right) c_{R\sigma^{\prime}}(\tau) + 
\mbox{h.c.}
\end{equation}
Here, $\tau$ denotes the time, $t$ is a hopping matrix element, $\lambda$ is a dimensionless constant, $\boldsymbol{\sigma} = 
(\sigma^x, \sigma^y, \sigma^z)$ is a vector formed by the three Pauli matrices, and $\boldsymbol{S}(\tau) = 
(S^x(\tau), S^y(\tau), S^z(\tau))$ denotes the spin of the magnetic impurity (for simplicity we consider here a single spin, 
but the theory can be readily generalized to the case of an arbitrary number of spins). The time-dependent phase accompanying 
the hopping is given by $\phi(\tau) = \omega_0 \tau + \alpha \sin (\omega_r \tau)$, where $\omega_0 = eV/\hbar$ ($V$ being the 
dc bias voltage), $\omega_r$ is the microwave frequency, and $\alpha = eV_\text{ac} / \hbar \omega_r$ ($V_\text{ac}$ being the 
ac bias voltage generated across the junction by the microwave field). 

This tunneling Hamiltonian incorporates the two basic ingredients in ESR-STM: the elastic and inelastic electron tunneling, 
and the microwave field. The elastic and inelastic electron tunneling is accounted for by the term in the brackets in Eq.~\eqref{eq-HT}. 
The $\delta$-function describes the regular elastic tunneling, while the second term accounts for the interaction with the spin 
system during which the tunneling electrons may lose/gain energy. The dimensionless constant $\lambda$ quantifies the relative 
contribution of this inelastic channel. In the absence of microwaves, Eq.~(\ref{eq-HT}) is the starting point of the established 
theory of single-spin inelastic tunneling spectroscopy \cite{fernandez-rossier_theory_2009,fransson_theory_2010,ternes_spin_2015}, 
which has been confirmed in numerous experiments \cite{ternes_spin_2015,Delgado_Fernandez-Rossier_2017,loth_controlling_2010}. 
On the other hand, we also describe the effect of the microwave field in terms of an ac bias voltage $V_\text{ac}$ in the junction. 
Here, we have used a standard unitary transformation to include the ac bias as a phase factor in the hopping matrix element 
\cite{cuevas2002subharmonic,cuevas2017molecular}. The parameter $\alpha$ describes the strength of the coupling to the electromagnetic 
field. This approach follows the well-established theory of photon-assisted tunneling \cite{tien_multiphoton_1963}, which has extremely 
successfully explained the physics of microwave-irradiated experiments in atomic-scale systems \cite{cuevas2017molecular,kohler2005driven,
roychowdhury_microwave_2015,kot_microwave-assisted_2020,peters_resonant_2020,siebrecht_microwave_2023}. We note that the approach that 
we are about to describe naturally reduces to the theory of single-spin inelastic tunneling spectroscopy in the absence of microwaves 
and to the photon-assisted tunneling theory in the absence of spin-flip tunneling. 

For later convenience we rewrite the tunneling Hamiltonian of Eq.~(\ref{eq-HT}) by separating the degrees of freedom related to the
electrodes and to the spin system as follows
\begin{eqnarray} \label{eq-HT-v2}
    H_\text{T}(\tau) & = & t e^{i \phi(\tau)} \sum_{\mu} \sum_{\sigma, \sigma^\prime} 
    \left[ c^\dagger_{L\sigma}(\tau) c_{R \sigma^\prime}(\tau) + c^\dagger_{R\sigma}(\tau) c_{L \sigma^\prime}(\tau) \right]  \nonumber \\ 
    & & \hspace{2.2cm} \Lambda^{\mu}_{\sigma \sigma^\prime} \otimes {\cal S}^{\mu}(\tau) ,   
\end{eqnarray}
where the index $\mu$ goes from $0$ to $3$, and we have defined
\begin{eqnarray}
	\boldsymbol{\Lambda}_{\sigma \sigma^\prime} & = & \Big( \delta_{\sigma \sigma^\prime}, \frac{\lambda}{2} \sigma^+_{\sigma \sigma^\prime}, 
 \frac{\lambda}{2} \sigma^-_{\sigma \sigma^\prime}, \lambda \sigma^z_{\sigma \sigma^\prime}\Big) , \label{eq-Lambda} \\
	\boldsymbol{\cal S}(\tau) & = & \Big(\mathbb{1}, \frac{1}{\hbar}S^{-}(\tau), \frac{1}{\hbar}S^{+}(\tau), \frac{1}{\hbar}S^{z}(\tau) \Big) , 
	\label{eq-S}
\end{eqnarray}
where $\sigma^{\pm}_{\sigma \sigma^\prime} = \sigma^x_{\sigma \sigma^\prime} \pm i \sigma^y_{\sigma \sigma^\prime}$ and 
$S^{\pm} = S^x \pm i S^y$.

\subsection{Tunneling current} \label{sec-tuncur}

Within our model, the current operator is given by \cite{cuevas2017molecular}
\begin{equation}
    \hat{I}(\tau) = \frac{ie}{\hbar} \sum_{\sigma} \left[ c^\dagger_{L \sigma} (\tau) c_{L \sigma} (\tau) ,  H_\text{T}(\tau) \right] .
\end{equation}
Using linear response theory \cite{fetter2012quantum} (i.e., first-order perturbation theory in $H_\text{T}$), the
expectation value of the time-dependent current can be expressed as
\begin{equation}
    I(\tau) = -\frac{i}{\hbar} \int_{-\infty}^{\tau} \braket{\left[ \hat{I}(\tau) , H_\text{T}(\tau^\prime) \right]}_0 d\tau^\prime,
    \label{eq: linear current}
\end{equation}
where $\braket{ \;}_0$ denotes an expectation value computed in the absence of the perturbation given by $H_\text{T}$. Making now use 
of Eq.~(\ref{eq-HT-v2}) and Wick's theorem \cite{fetter2012quantum}, it can be shown that the time-dependent current in the presence 
of microwaves is given by
\begin{widetext}
\begin{eqnarray} \label{eq-Itau}
    I(\tau) & = & \frac{2e}{h} \sum_{n, m = -\infty}^{\infty} J_{m}(\alpha) J_{n+m}(\alpha) \sum_{\alpha, \beta}
    \int_{-\infty}^{\infty} d\omega \int_{-\infty}^{\infty} d\omega^\prime \int_{0}^{\infty} \frac{d\tau^\prime}{h} \\ & &
     \Re \left\lbrace  \left[  e^{in\omega_r \tau} e^{i(\omega - \omega^\prime)\tau^\prime} 
     \vec{\Gamma}_{\alpha \beta}(\omega, \omega^\prime + \omega_0 + m \omega_r) -  
     e^{-in\omega_r \tau} e^{-i(\omega - \omega^\prime)\tau^\prime} 
     \cev{\Gamma}_{\alpha \beta}(\omega, \omega^\prime + \omega_0 + m \omega_r) \right]  
     \braket{ {\cal S}^{\alpha \dagger}(\tau) {\cal S}^{\beta}(\tau-\tau^\prime) }_0 
    \right\rbrace, \nonumber
\end{eqnarray}
\end{widetext}
Here, $J_n(\alpha)$ is a Bessel function of the first kind of order $n$ and $\braket{A}_0 = \mathrm{Tr} \left[ \rho(\tau) A \right]$, 
with $\rho(\tau)$ being the reduced density operator of the spin system which will be determined below. Moreover, we have introduced 
the spectral rates
\begin{eqnarray}
    \vec{\Gamma}_{\alpha \beta} (\omega, \omega^\prime) & = & t^2 \mathrm{Tr} \left[  g^{+-}_L(\omega) \Lambda^{\alpha \dagger} 
    g^{-+}_R(\omega^\prime) \Lambda^{\beta}  \right] , \label{eq-Gamma-LR} \\
    \cev{\Gamma}_{\alpha \beta} (\omega, \omega^\prime) & = & t^2 \mathrm{Tr} \left[  g^{-+}_L(\omega) \Lambda^{\beta} 
    g^{+-}_R(\omega^\prime) \Lambda^{\alpha \dagger}  \right] \label{eq-Gamma-RL} ,
\end{eqnarray}
associated with electron transitions from the left electrode to the right one and vice versa. The same spectral rates will also later be used to calculate the dissipator (cf.\ Eq.\ \eqref{eq-dissipator}) in the equation of motion for the density matrix $\rho(\tau)$. These functions are expressed 
in terms of the lead Green's functions defined as
\begin{eqnarray}
    g^{+-}_{j \sigma \sigma^\prime}(\omega) & = & i \int_{-\infty}^{\infty}  e^{i \omega \tau} \braket{c^{\dagger}_{j \sigma^\prime}(0) 
    c_{j \sigma}(\tau)}_0 d\tau, \label{eq-g+-} \\
    g^{-+}_{j \sigma \sigma^\prime}(\omega) & = & -i \int_{-\infty}^{\infty} e^{i \omega \tau} \braket{c_{j \sigma}(\tau) 
    c^{\dagger}_{j \sigma^\prime}(0)}_0 d\tau . \label{eq-g-+}
\end{eqnarray}
where $j = L,R$. The Green's functions describe the electronic structure of the electrodes and also contain the information of the 
occupation factors. Finally, the trace $\mathrm{Tr}$ appearing in Eqs.~\eqref{eq-Gamma-LR} and \eqref{eq-Gamma-RL} refers to the 
$2 \times 2$ spin space. Thus, the current is expressed in terms of the basic properties of the electrodes via their Green's functions, 
and spin correlation functions of the magnetic system. The Green's functions will be specified in more detail below when we discuss 
the comparison with the experiments. The calculation of the correlation functions of the spin system will be detailed in the next subsection.

\subsection{Dynamics of a general spin system}

To complete the calculation of the current we need to compute the spin correlation functions $\braket{ {\cal S}^{\alpha \dagger}(\tau) {\cal S}^{\beta}(\tau-\tau^\prime) }_0 $ entering Eq.~(\ref{eq-Itau}). To do so, 
we determine the reduced density matrix of the spin system $\rho(\tau)$ by employing standard techniques of the theory of open quantum 
systems. There are different approaches to obtain the quantum master equation describing the dynamics of the reduced density matrix, 
such as the Bloch-Redfield equations \cite{cohen1992basic}. We find it more convenient to use the Lindblad approach and we shall follow 
closely Ref.~\onlinecite{breuer_open_2002}. Within the Lindblad approach, $\rho(\tau)$ satisfies the following equation of motion in the
interaction picture \cite{breuer_open_2002} 
\begin{equation}
    \frac{d \rho(\tau)}{d \tau} = -\frac{i}{\hbar} \left[H_\text{r}(\tau) + H_\mathrm{LS} , \rho(\tau) \right] + \mathcal{D} 
    \left[ \rho(\tau) \right],
\end{equation}
where $H_\text{r}(\tau)$ describes the interaction with the microwave (specified in more detail below), $H_\mathrm{LS}$ is the Lamb 
shift Hamiltonian, and $\mathcal{D}\left[ \rho(\tau) \right]$ is the dissipator operator. The expressions for $H_\mathrm{LS}$ and 
$\mathcal{D}\left[ \rho(\tau) \right]$ depend on the system-bath interaction $H_\text{I}$. To proceed, we shall assume that $H_\text{I}$ 
can be factorized as follows
\begin{equation}
	H_\text{I} = \sum_\alpha A_\alpha \otimes B_\alpha,
\end{equation}
where $A$ is an operator related to the spin system and $B$ is a bath operator. With this decomposition, the dissipator and the Lamb 
shift Hamiltonian are given by \cite{breuer_open_2002}  
\begin{eqnarray}
    \mathcal{D} \left[ \rho(\tau) \right] & = & \frac{1}{\hbar^2} \sum_{\omega} \sum_{\alpha \beta} \gamma_{\alpha \beta} (\omega) 
    \left[ A_{\beta}(\omega) \rho(\tau) A^{\dagger}_{\alpha}(\omega) - \right. \nonumber \\ & &
    \left. - \frac{1}{2} \left\lbrace A_{\alpha}^{\dagger}(\omega) A_{\beta}(\omega), \rho(\tau)\right\rbrace \right], \label{eq-dissipator}
\end{eqnarray}
\begin{equation}
    \mathcal{H}_{\mathrm{LS}} = \frac{1}{\hbar} \sum_\omega \sum_{\alpha \beta} h_{\alpha \beta} (\omega) 
    A^{\dagger}_{\alpha}(\omega) A_{\beta}(\omega),
    \label{eq-ls}
\end{equation}
where
\begin{align}
     \frac{1}{2} \gamma_{\alpha \beta} (\omega) + i h_{\alpha \beta} (\omega) = \int_{0}^{\infty} d\tau e^{i\omega\tau} 
     \braket{B^\dagger_\alpha (\tau) B_\beta (0)}.
\end{align}
In Eqs.\ \eqref{eq-dissipator} and \eqref{eq-ls}, we have defined the operator
\begin{equation}
    A_{\alpha}(\omega) = \sum\limits_{\varepsilon'-\varepsilon = \omega} \mathcal{P}(\varepsilon') A_{\alpha}
    \mathcal{P}(\varepsilon).
\end{equation}
Here, $\mathcal{P}(\varepsilon)$ is a projection operator onto the eigenstate of $H_\text{S}$ belonging to the eigenvalue
$\varepsilon$ and the sum is over all pairs of states with energy $\varepsilon$ and $\varepsilon'$ having the energy difference 
$\omega = \varepsilon' - \varepsilon$.

There are three natural baths that can contribute to the relaxation and decoherence of the spin system: the tunneling electrons, 
the microwave field, and the substrate. For simplicity, we shall assume that the action of these different baths is additive 
(i.e., they are independent). We include the impact of the tunneling electrons on the spin dynamics using the general formalism 
above. In this case, the system-bath interaction $H_\text{I}$ is given by $H_\text{T}$ in Eq.~\eqref{eq-HT-v2} without microwaves 
(i.e.\ $\alpha = 0$). Thus, we find that $A_\alpha = {\cal S}^\alpha$, see Eq.~\eqref{eq-S}, and $B_\alpha(\tau) = t 
\sum_{\sigma, \sigma^\prime} \left[ c^\dagger_{L\sigma}(\tau) c_{R \sigma^\prime}(\tau) + c^\dagger_{R\sigma}(\tau) 
c_{L \sigma^\prime}(\tau)  \right] \Lambda^{\alpha}_{\sigma \sigma^\prime}$, where the index $\alpha$ runs from $0$ to $3$ 
(cf.\ Eq.\ \eqref{eq-Lambda}). The quantities $\gamma_{\alpha \beta} (\omega)$ and $h_{\alpha \beta}(\omega)$ can be expressed 
in terms of the spectral rates of Eqs.~\eqref{eq-Gamma-LR} and \eqref{eq-Gamma-RL} as follows
\begin{eqnarray} 
    \gamma_{\alpha \beta}(\omega) & = & \int_{-\infty}^{\infty} \frac{d\omega_1}{2 \pi} \left\lbrace 
    \vec{\Gamma}_{\alpha \beta}(\omega_1, \omega_1 + \omega_0 + \omega) + \right. \nonumber \\ & & \left. \hspace*{1.5cm}
    \cev{\Gamma}_{\alpha \beta}(\omega_1, \omega_1 + \omega_0 - \omega) \right\rbrace ,  \label{eq-gamma} \\
    h_{\alpha \beta}(\omega) & = & \int_{-\infty}^{\infty} \frac{d\omega_1}{2 \pi} \int_{-\infty}^{\infty} 
    \frac{d\omega_2}{2 \pi} \left\lbrace \frac{\vec{\Gamma}_{\alpha \beta}(\omega_1, \omega_2)}{\omega + \omega_2 - \omega_1 - \omega_0} + 
    \right. \nonumber \\ & & \left. \hspace*{2.65cm}
    \frac{\cev{\Gamma}_{\alpha \beta}(\omega_1, \omega_2)}{\omega - \omega_2 + \omega_1 + \omega_0} \right\rbrace. \label{eq-h}
\end{eqnarray}

The interaction between the spin system and the substrate can be treated essentially in the same way as the tunneling 
electrons, while the impact of the microwave field (in the form of spontaneous emission) is a textbook result that we shall
simply borrow here \cite{cohen1992basic}. We shall be more specific on how we model these two interactions in our discussion 
of a \spinhalf\ system in Section~\ref{sec-TLS-spin}. 

\section{Application to a \spinhalf\ system} \label{sec-TLS}

Let us now apply the general theory described in Section~\ref{sec-general} to the case of a \spinhalf\ system. Upon application 
of a static magnetic field, the two degenerate spin levels are Zeeman-split and this two-level system can be described by the 
Hamiltonian: $H_\text{S} = \hbar \omega_a \ket{a}\bra{a} + \hbar \omega_b \ket{b}\bra{b}$ where state $\ket{a}$ is the spin-up 
state (or ground state) and $\ket{b}$ is the spin-down state (or excited state). This Hamiltonian can be written in a standard 
matrix form as $H_\text{S} = \frac{1}{2} \hbar \omega_{ba} \sigma^z$, where $\omega_{ba} = \omega_b - \omega_a > 0$. The 
interaction with the radiation field is given by
\begin{equation}
    H_\text{r}(\tau) = -\hbar \Omega \cos(\omega_r \tau) \sigma^x,
\end{equation}
where $\Omega$ is the Rabi frequency carrying the information of the intensity of the microwave drive. At this point, we do not need to further 
specify the origin of the Rabi frequency. However, it must be related to the ac magnetic field in the junction region, as we discuss in Sec.~\ref{sec-EM}. In the following, we shall discuss the results of the general theory applied to this two-level system.

\subsection{Spin dynamics} \label{sec-TLS-spin}

The density matrix of a \spinhalf\ system is a $2\times2$ matrix. For a pure system, it is defined as $\rho = \ket{\psi} \bra{\psi}$, 
where $\ket{\psi}$ is the wave function of the two-level system in the basis $\left\lbrace \ket{a}, \ket{b} \right\rbrace$. For this system, 
the dissipator and the Lamb shift operators adopt the form
\begin{eqnarray}
   \mathcal{D}\left[ \rho \right] & = & \Gamma_{a \rightarrow b}
    \begin{pmatrix} -\rho_{aa} & \rho_{ab}/2 \\ \rho_{ba}/2 & \rho_{bb} \end{pmatrix}
    + \Gamma_{b \rightarrow a} \begin{pmatrix} \rho_{aa} & \rho_{ab}/2 \\ \rho_{ba} & -\rho_{bb} \end{pmatrix} - \nonumber \\
    & & - \Gamma_{ab}^{\mathrm{ad}} \begin{pmatrix} 0 & \rho_{ab} \\ \rho_{ba} & 0 \end{pmatrix} ,  \\
   \mathcal{H}_{\mathrm{LS}} & = & \frac{\hbar \Delta_{ab}}{2} \begin{pmatrix} 1 & 0 \\ 0 & -1 \end{pmatrix}.
\end{eqnarray}
Here, $\Gamma_{a \rightarrow b} \equiv \gamma_{11} (-\omega_{ba}) / \hbar^2$ is the transition rate from $a$ to $b$, 
$\Gamma_{b \rightarrow a} \equiv \gamma_{22} (\omega_{ba}) / \hbar^2 + \Gamma_0$ is the transition rate from $b$ to $a$, 
$\Gamma^{\mathrm{ad}}_{ab} \equiv \gamma_{33} (0)/2 \hbar^2 + \gamma_0$ is the pure decoherence rate. The renormalization energy $\Delta_{ab} = \Delta_{ab}^{\mathrm{ad}} + \Delta_{ab}^{\mathrm{nonad}}$ consists of two contributions: an adiabatic energy renormalization $\Delta^{\mathrm{ad}}_{ab} = 
h_{03}(0) / \hbar^2$ and a nonadiabatic energy renormalization $\Delta^{\mathrm{nonad}}_{ab} = \left[ h_{11}(-\omega_{ba}) - 
h_{22}(\omega_{ba}) \right] / \hbar^2 $. Explicit expressions for these 
parameters follow from Eqs.~\eqref{eq-gamma}-\eqref{eq-h} and are given by
\begin{widetext}
\begin{align}
\begin{split} \label{eq-Gamma-ab}
    \Gamma_{a\rightarrow b} = \left(\frac{ \lambda t  }{\hbar} \right)^2 \int_{-\infty}^{\infty} \frac{d\omega}{2\pi}  
    \left[ g^{+-}_{L\downarrow}(\omega) g^{-+}_{R\uparrow}(\omega + \omega_0 - \omega_{ba}) + g^{-+}_{L\uparrow}(\omega) 
    g^{+-}_{R\downarrow}(\omega + \omega_0 + \omega_{ba}) \right],
\end{split}
\end{align}
\begin{align}
\begin{split} \label{eq-Gamma-ba}
    \Gamma_{b\rightarrow a} = \left(\frac{ \lambda t }{\hbar} \right)^2 \int_{-\infty}^{\infty} \frac{d\omega}{2\pi}  
    \left[ g^{+-}_{L\uparrow}(\omega) g^{-+}_{R\downarrow}(\omega + \omega_0 + \omega_{ba}) + g^{-+}_{L\downarrow}(\omega)
    g^{+-}_{R\uparrow}(\omega + \omega_0 - \omega_{ba}) \right] + \Gamma_0,
\end{split}
\end{align}
\begin{align}
\begin{split} \label{eq-Gamma-ad}    
    \Gamma^{\mathrm{ad}}_{ab} = \frac{1}{2} \left( \frac{ \lambda t}{\hbar} \right)^2 \int_{-\infty}^{\infty} \frac{d\omega}{2\pi}  
    \left[ g^{+-}_{L\uparrow}(\omega)g^{-+}_{R\uparrow}(\omega + \omega_0) +  
    g^{-+}_{L\uparrow}(\omega)g^{+-}_{R\uparrow}(\omega + \omega_0) + 
    g^{+-}_{L\downarrow}(\omega)g^{-+}_{R\downarrow}(\omega + \omega_0) + 
    g^{-+}_{L\downarrow}(\omega)g^{+-}_{R\downarrow}(\omega + \omega_0) \right] + \gamma_0,
\end{split}
\end{align}
\begin{align}
    \Delta^{\mathrm{ad}}_{ab} = \frac{2\lambda t^2 }{\hbar^2} \int_{-\infty}^{\infty} \frac{d\omega}{2\pi} \int_{-\infty}^{\infty} 
    \frac{d\omega^\prime}{2\pi} 
    \frac{g^{-+}_{L\uparrow}(\omega)g^{+-}_{R\uparrow}(\omega^{\prime}) + 
    g^{+-}_{L\downarrow}(\omega)g^{-+}_{R\downarrow}(\omega^{\prime}) - 
    g^{-+}_{L\downarrow}(\omega)g^{+-}_{R\downarrow}(\omega^{\prime}) - 
    g^{+-}_{L\uparrow}(\omega)g^{-+}_{R\uparrow}(\omega^{\prime})}{\omega^\prime- \omega - 
    \omega_0 }
    \label{eq: ad shift}
\end{align}
\begin{align}
    \Delta^{\mathrm{nonad}}_{ab} = \left( \frac{ \lambda t}{\hbar} \right)^2 \int_{-\infty}^{\infty} \frac{d\omega}{2\pi} 
    \int_{-\infty}^{\infty} \frac{d\omega^\prime}{2\pi} \left\lbrace \frac{g^{-+}_{L\uparrow}(\omega)g^{+-}_{R\downarrow}(\omega^{\prime}) +
     g^{+-}_{L\uparrow}(\omega)g^{-+}_{R\downarrow}(\omega^{\prime})}{\omega^\prime - \omega - \omega_0  - \omega_{ba}} - 
     \frac{g^{-+}_{L\downarrow}(\omega)g^{+-}_{R\uparrow}(\omega^{\prime}) + g^{+-}_{L\downarrow}(\omega)g^{-+}_{R\uparrow}(\omega^{\prime})}
     {\omega^\prime - \omega - \omega_0  + \omega_{ba}} \right\rbrace,
    \label{eq: nonad shift}
\end{align}
\end{widetext}
The parameters $\Gamma_0$ and $\gamma_0$ are introduced here as phenomenological decay and decoherence rates, respectively, that carry contributions from both the substrate coupling and spontaneous emission. We have assumed that the lead Green's functions are diagonal 
in spin space (i.e.\ no spin mixing in the electrodes). Also, the Lamb shift may be absorbed in the bare frequency $\omega_{ab}$, 
as it is usually done in different contexts \cite{Dung2002_resonant}. However, the Lamb shift may be nonnegligible affecting the 
determination of the value for the tip magnetic field. In addition, the previous expressions are analyzed in Appendix~\ref{sec-app1} 
in relevant limiting cases to give analytical insight.

In a rotating frame (rotating with the same frequency as the microwave enabling us to use the rotating-wave approximation where 
$\omega_{ba} \sim \omega_r$), the quantum master equation for the density matrix in terms of the quantities defined above adopts the form
\begin{eqnarray} \label{eq-QTE}
    \frac{d \rho_{aa}(\tau)}{d\tau} & = & -\frac{i \Omega}{2} \left[ \tilde{\rho}_{ba}(\tau) - \tilde{\rho}_{ab}(\tau) \right] - 
    \Gamma_{a \rightarrow b} \rho_{aa}(\tau) + \nonumber \\ & & \Gamma_{b \rightarrow a} \rho_{bb}(\tau), \\
    \frac{d \rho_{bb}(\tau)}{d\tau} & = & \frac{i \Omega}{2} \left[ \tilde{\rho}_{ba}(\tau) - \tilde{\rho}_{ab}(\tau) \right] - 
    \Gamma_{b \rightarrow a} \rho_{bb} (\tau) + \nonumber \\ & & \Gamma_{a \rightarrow b} \rho_{aa}(\tau), \\
    \frac{d \tilde{\rho}_{ab}(\tau)}{d\tau} & = & -\frac{i \Omega}{2} \left[ \rho_{bb}(\tau) - \rho_{aa}(\tau) \right] - 
    \nonumber \\ & & (\gamma + i \delta) \tilde{\rho}_{ab} (\tau) .
\end{eqnarray}
Here, $\tilde{\rho}_{ba} = \tilde{\rho}_{ab}^{*}$, $\rho_{ab}(\tau) = e^{i(\omega_r - \omega_{ba})\tau} \tilde{\rho}_{ab}$ is the 
relation between the rotating frame and the interaction picture density matrix. Furthermore, $\gamma$ is the effective decoherence 
rate defined as
\begin{equation}
    \gamma = \Gamma^{\mathrm{ad}}_{ab} + \left( \Gamma_{b \rightarrow a} + \Gamma_{a \rightarrow b} \right)/2,
\end{equation}
and $\delta = \omega_r - \omega_{ba} + \Delta_{ab}$ is the detuning of the microwaves (including the second order perturbation to 
the energy gap $\Delta_{ab}$).

In this work, we focus on the case of continuous illumination, where in the long time limit, we reach a stationary situation for 
$\rho_{aa}$, $\rho_{bb}$ and $\tilde{\rho}_{ab}$. The stationary solution for this rotating frame is given by
\begin{eqnarray} 
    \tilde{\rho}(\tau \rightarrow \infty) & = & \frac{1}{2(\delta^2 + \gamma^2) \Gamma + 
    2 \gamma \Omega^2} \times \nonumber \\ & & \hspace*{-2cm}
    \begin{pmatrix}
        \gamma \Omega^2 + 2\Gamma_{b \rightarrow a} (\delta^2 + \gamma^2) &  \Omega \Gamma^{\prime} (\delta + i\gamma) \\
        \Omega \Gamma^{\prime} (\delta - i\gamma) & 
        \gamma \Omega^2 + 2\Gamma_{a \rightarrow b} (\delta^2 + \gamma^2)
    \end{pmatrix} , \label{eq-rho-str}
\end{eqnarray}
where $\rho_{aa,bb} (\tau) = \tilde{\rho}_{aa,bb}$, $\rho_{ab} (\tau) = e^{i (\omega_r - \omega_{ba}) \tau} \tilde{\rho}_{ab}$. 
We also used the abbreviation
\begin{equation}
    \Gamma = \Gamma_{b \rightarrow a} + \Gamma_{a \rightarrow b}\text{\quad and\quad}\Gamma^{\prime} = \Gamma_{b \rightarrow a} - 
    \Gamma_{a \rightarrow b}.
\end{equation}
The solution for the density matrix in Eq.~\eqref{eq-rho-str} can be used in Eq.~\eqref{eq-Itau} for the calculation
of the dc current, which will be done in the next subsection.

With knowledge of the decay and decoherence rates we can determine the characteristic time scales for relaxation and decoherence
of the spin system. Following Refs.~\cite{Delgado_Lorente_2021,Delgado_Fernandez-Rossier_2017}, we define the following time scales:
\begin{equation} \label{eq-times}
  T_1 = \frac{1}{\Gamma},\quad 
  T_2^{*} = \frac{1}{ \Gamma^{\mathrm{ad}}_{ab} }, \quad
  T_2 = \frac{1}{\gamma},
\end{equation}
where $T_1$ represents the relaxation time, $T_2^{*}$ the pure decoherence time, and $T_2$ the total decoherence time. We note 
that all of these times depend on the applied bias voltage and correspondingly on the tunneling current through 
Eqs.\ \eqref{eq-Gamma-ab} to \eqref{eq-Gamma-ad}.

\subsection{dc current} \label{sec-TLS-current}

We are now in position to provide the key results for the current in a \spinhalf\ system. We shall focus on the dc current and consider 
first the special case, in which the spin of the magnetic impurity is parallel to the spin quantization axis in the STM tip (due to 
a high static magnetic field that aligns all spins in the system). In this case, the dc current adopts a very appealing form given by
\begin{equation} \label{eq-TG}
    I(V, \alpha) = \sum_{n = -\infty}^{\infty} J^2_n(\alpha) \, I^{\mathrm{dark}}(V + n \hbar \omega_r/e) \, ,
\end{equation}
where $I^{\mathrm{dark}}(V)$ is formally the current in absence of microwaves, but all the properties of the spin system have to be computed 
taking into account the effect of the microwave field. Notice that Eq.~\eqref{eq-TG} has the form of the Tien-Gordon formula that is well-known 
in the context of photon-assisted tunneling \cite{tien_multiphoton_1963,cuevas2017molecular,kohler2005driven}. The dark current is given by 
the sum of three contributions
\begin{equation} \label{eq-Idark}
    I^{\mathrm{dark}}(V) = I_{\mathrm{el}}(V) + I_{\mathrm{int}}(V) + I_{\mathrm{inel}}(V) ,
\end{equation}
which are given in terms of the electrode Green's functions and the elements of the stationary density matrix of Eq.~\eqref{eq-rho-str}
as follows 
\begin{widetext}
\begin{eqnarray}
 I_{\mathrm{el}} (V) & = &  \frac{4 \pi^2 e t^2 }{h} \sum_\sigma \int_{-\infty}^{\infty} \varrho_{L \sigma}(E-eV) \varrho_{R \sigma}(E) 
 \left[f(E-eV)- f(E) \right] dE = I^{\uparrow}_{\mathrm{el}}(V) + I^{\downarrow}_{\mathrm{el}}(V) , \label{eq-Iel} \\
 I_{\mathrm{int}} (V) & = & \frac{2 \lambda}{\hbar} \braket{S^z} \left[ I^{\uparrow}_{\mathrm{el}}(V) - 
 I^{\downarrow}_{\mathrm{el}}(V)\right] = \lambda \left[ I^{\uparrow}_{\mathrm{el}}(V) - 
 I^{\downarrow}_{\mathrm{el}}(V)\right] \left( \rho_{aa} - \rho_{bb} \right)  , \label{eq-Iint} \\
 I_{\mathrm{inel}}(V) & = & \frac{\lambda^2}{4} I_{\mathrm{el}}(V) - \frac{\lambda^2}{2} \left[ I_{+}(V)+I_{-}(V) \right]
+ \frac{\lambda^2}{2} \left[I_{+}(V)-I_{-}(V)\right] \left( \rho_{aa} - \rho_{bb} \right) , \label{eq-Iinel}
\end{eqnarray}
with the definitions
\begin{align}
    I_{+}(V) & = \frac{e t^2}{h} \int_{-\infty}^{\infty} 
    \left[g^{+-}_{L\downarrow}(E) g^{-+}_{R\uparrow}(E+eV-\hbar \omega_{ba})  - 
    g^{-+}_{L\uparrow}(E) g^{+-}_{R\downarrow}(E+eV+\hbar \omega_{ba}) \right]  dE , \nonumber \\
    I_{-}(V) & = \frac{e t^2}{h} \int_{-\infty}^{\infty} 
    \left[ g^{+-}_{L\uparrow}(E) g^{-+}_{R\downarrow}(E+eV+\hbar \omega_{ba}) - 
    g^{-+}_{L\downarrow}(E) g^{+-}_{R\uparrow}(E+eV-\hbar \omega_{ba}) \right]  dE . \nonumber
\end{align}
\end{widetext}
Here, $\varrho_{j \sigma}(E)$ is the density of states (DOS) of electrode $j=L,R$ and $f(E)$ is the Fermi function. The first
contribution $I_{\mathrm{el}}(V)$ is the elastic current flowing through the junction, which is given by the standard
tunneling formula. The third contribution $I_{\mathrm{inel}}(V)$ is the inelastic current that results from the spin-flip processes 
undergone by the tunneling electrons. The second contribution $I_{\mathrm{int}}(V)$, which we refer to as interference current, can 
be seen as resulting from the interference between the two previous processes. The second term is expected to dominate the signal 
involving the spin system because it is first order in $\lambda$ (this is confirmed below by our comparison with the experiments). 
This interference term is proportional to both the spin polarization of the spin system along the $z$-direction $\braket{S^z} = 
\frac{\hbar}{2} (\rho_{aa} - \rho_{bb})$ and to the current polarization $I_\text{pol}(V) \equiv I^{\uparrow}_{\mathrm{el}}(V) - 
I^{\downarrow}_{\mathrm{el}}(V)$ in the absence of microwaves. Therefore, this contribution is only finite when there is a population 
difference ($\rho_{aa} \neq \rho_{bb} $) and moreover the elastic current has a finite spin polarization. This explains the need for 
a spin-polarized tip to observe an ESR signal and shows that \emph{ESR spectra are a measure of the change in magnetoresistance that 
results from the resonant transitions between the electronic levels induced by the microwave field.} It is also worth mentioning that 
the inelastic term of Eq.~\eqref{eq-Iinel} also contributes to the ESR signal, which requires a finite current polarization as well.
The contribution of the interference and inelastic terms to the ESR spectrum can be computed combining Eqs.~\eqref{eq-Iint}
and \eqref{eq-Iinel} with Eq.~\eqref{eq-TG}. Defining the ESR lineshape by subtracting the off-resonant current (for $\delta \to 
\infty$), we arrive at the following expression describing the ESR signal in terms of the relaxation and decoherence times
(cf.\ Eq.~\eqref{eq-times})
\begin{equation} 
    \mbox{ESR}(\delta) = I_\text{sat} \left( \frac{\Omega^2 T_1 T_2}{\delta^2 T^2_2 + 1 + \Omega^2 T_1 T_2} \right) ,  \label{eq-ESR}
\end{equation}
where we have defined
\begin{eqnarray} \label{eq-Isat} 
    I_\text{sat} & = & - \lambda \sum_n J^2_n(\alpha) I_\text{pol} (V + n \hbar \omega_r/e) - \\
    & & \frac{\lambda^2}{2} \sum_n J^2_n(\alpha) \left[ I_{+} (V + n \hbar \omega_r/e) - 
    I_{-} (V + n \hbar \omega_r/e) \right] \nonumber .
\end{eqnarray}
Equation~\eqref{eq-ESR} predicts an ESR peak height given by
\begin{equation} \label{eq-Ipeak}
I_\text{peak}  = I_\text{sat} \; \frac{\Omega^2 T_1 T_2}{1 + \Omega^2 T_1 T_2} ,
\end{equation}
where $I_\text{sat}$ is the saturation value of the current peak when $\Omega \to \infty$. The corresponding linewidth 
is given by ${\cal W} = \frac{1}{T_2} \sqrt{ 1 + \Omega^2 T_1 T_2}$. These results coincide with the phenomenological formulas 
that have been employed to describe the experimental ESR spectra, see e.g., Refs.~\cite{baumann_electron_2015,willke_probing_2018}. 
Thus, our derivation here can be seen as a rigorous justification of those heuristic formulas. We point out here that the time 
scales $T_1$ and $T_2$ in these formulas are not phenomelogical parameters, but they are given by the expressions of 
Eqs.~\eqref{eq-times} and (\ref{eq-Gamma-ab}-\ref{eq-Gamma-ad}) which make clear predictions about their dependence on the bias 
voltage and the tip-sample distance (encoded in the hopping parameter $t$). Moreover, Eq.~(\ref{eq-Isat}) contains the information 
about the degree of spin-polarization of a STM tip to obtain sizeable ESR signals. In Appendix~\ref{sec-app1}, we provide additional 
analytical insight into these results.

The current in the absence of microwaves can be obtained by setting $\alpha = 0$ in the equations above. In that case, the current 
is given by the dark current in Eqs.~\eqref{eq-Idark}-\eqref{eq-Iinel}, where the density matrix elements $\rho_{aa}$ and $\rho_{bb}$ 
are computed in the absence of microwaves. Furthermore, the result for the current coincides with the results reported in the literature 
in the absence of microwaves \cite{fernandez-rossier_theory_2009,fransson_theory_2010,ternes_spin_2015}. Lastly, the division into 
three current contributions above is not generic of a \spinhalf\ system, but applies to any spin system.

\subsection{Homodyne detection} \label{sec-TLS-homodyne}

In the previous section, we have assumed that the spins of the system and the tip polarization are parallel. This led 
us to a result for the ESR lineshape summarized in Eq.~(\ref{eq-ESR}) that predicts symmetric spectra with respect to the 
detuning $\delta$. However, asymmetric spectra have been reported in the literature \cite{bae_enhanced_2018,yang2019tuning,willke_probing_2018}, 
in particular close to zero bias voltage, which we will discuss below. In the following, we assume that the asymmetry arises 
when the static magnetic field does not completely align the tip and impurity spin. Other proposals trace the origin of the asymmetry 
back to the failure of the rotating wave approximation \cite{shavit_generalized_2019}. Focusing on the two-level system, if the impurity 
spin and the magnetization of the tip are not collinear and form an angle $\theta \neq 0$, one can show that this induces a Larmor 
precession of the impurity magnetization, i.e.\ when $\omega_r \approx \omega_{ba}$, this leads to an expectation value for 
$S^z(\tau)$ of the form
\begin{eqnarray} \label{eq-homodyne1}
    \braket{S^z(\tau)} & = & \frac{\hbar \cos \theta }{2} \left[ \rho_{aa}(\delta) - \rho_{bb}(\delta) \right] + \\
    & & \frac{\hbar \sin \theta }{2} \left[ \tilde{\rho}_{ba} (\delta) e^{-i\omega_r \tau} + 
    \tilde{\rho}_{ab} (\delta) e^{i \omega_r \tau}  \right]. \nonumber
\end{eqnarray}
The second term here is time-dependent, which is determined by the coherences (i.e.\ off-diagonal elements) in the reduced density 
matrix. This time-dependent magnetization couples to the microwave field to give an additional contribution
to the dc current. This effect has been referred to as \emph{homodyne detection} \cite{willke_probing_2018}, in analogy
to what has been observed in the context of electric-field induced ferromagnetic resonance excitation \cite{nozaki2012electric}.
The dc current in this case can be obtained by inserting Eq.~\eqref{eq-homodyne1} into Eq.~\eqref{eq-Itau}. The main effect 
of the homodyne detection is the modification of the interference term, whose contribution to the dc current in the presence of 
microwaves becomes  
\begin{eqnarray} \label{eq-homodyne2}
	I_{\mathrm{int}}(V, \alpha) & = & \lambda \sum_{n = -\infty}^{\infty} J^2_{n}(\alpha) \,
	I_\text{pol} (V + n \hbar \omega_r/e) \times \\ & & \left[ \left( \rho_{aa} - \rho_{bb} \right) \cos{\theta} + 
	\frac{2 n}{\alpha} \Re \left\{ \tilde{\rho}_{ba} \right\} \sin{\theta} \right] . \nonumber
\end{eqnarray}
The first term inside the brackets gives the contribution that we have discussed in Section~\ref{sec-TLS-current}, but it is 
rescaled by $\cos \theta$. The second term is proportional to $\sin \theta$ and describes the new contribution that depends on the 
real part of the coherence, which is odd in $\delta$. As we shall show below where we analyze our own experimental results, this 
term is responsible for the asymmetry in the ESR spectra and explains the unconventional spectral lineshapes near zero bias voltage.

\section{Experimental results} \label{sec-exp}

With the theory described in previous sections, we can now not only quantitatively analyze the measured ESR signal, but also 
understand this signal in relation to the background signal. We can also describe the tunneling current with and without microwaves, 
which carries important information that allows us to independently determine the model parameters. Additionally, we can use
this theory to analyze the ESR signal as a function of applied magnetic field and/or microwave frequency, and also as 
function of applied bias voltage, which provides a stringent test of the theory. In this section, we shall briefly discuss
the experiments that were carried out to test the theory developed in this work.

As a testbed system for our theory we measured the ESR signal of a TiH molecule on MgO/Ag(100) as function of magnetic field and 
applied bias voltage, which is summarized in Fig.~\ref{fig:Fig1}(c). The TiH molecule forms a \spinhalf\ system on the MgO surface 
as has been observed in a number of different experiments before 
\cite{yang_coherent_2019,willke2018hyperfine,yang_engineering_2017,steinbrecher2021quantifying,kot_electric_2022}. The microwave 
is fed through an antenna, which radiates towards the tunnel junction through vacuum \cite{drost_combining_2022}. The microwave amplitude 
and frequency were kept constant during the measurements, while the magnetic field was swept. The data was acquired by chopping the 
microwave (on/off) at a frequency of 107\,Hz and detecting the signal in the tunneling current using a lock-in amplifier at that 
frequency \cite{paul_generation_2016,natterer_upgrade_2019,weerdenburg_scanning_2021}. The resulting measured current $\Delta I$ is 
effectively the difference between the tunneling current while the microwave is on $I_\text{on}$ and the tunneling current while the 
microwave is off $I_\text{on}$,
\begin{equation}
    \Delta I = I_\text{on} - I_\text{off}.
\end{equation}
Therefore, we have to calculate the tunneling current with and without the microwave radiating into the tunnel junction. Due to 
the sinusoidal sampling of the lock-in amplifier, the measured ESR signal will be smaller by a factor of $\pi/2$ compared to the 
calculated difference of the tunneling current with the microwave on and off. This has to be taken into account for a quantitative 
analysis. We will do so implicitly in the following. Further details on the experimental setup and measurements can be found in 
Appendix~\ref{sec-Methods}.

As one can see in Fig.~\ref{fig:Fig1}(c), there is a clear ESR peak whose height and width evolves with the bias voltage in a
non-trivial manner, which the theory must explain. Notice also that there is linear dependence of the ESR peak position on the
the bias voltage. We attribute this dependence to spin-electric coupling in the tunnel junction \cite{kot_electric_2022}, 
which we will only take into account phenomenologically here. On the other hand, the horizontal stripes are due to the tunneling 
electrons interacting with the microwave, which is not directly related to the ESR signal. In Fig.~\ref{fig:Fig1}(d), the 
corresponding calculated spectra using the theory of Sec.~\ref{sec-TLS} are shown for the same parameter range as in 
Fig.~\ref{fig:Fig1}(c), which are in good quantitative agreement with the experimental data. The details of how we calculated the 
theoretical spectra will be described in detail in the following section.

\section{Comparison with the experimental results} \label{sec-comp}

In this section we apply the results of the theory for a \spinhalf\ system discussed in Sec.~\ref{sec-TLS} to the analysis
of our experimental results. This analysis proceeds in two steps. First, we fit the tunneling current in the absence of 
microwaves, which allows us to determine most of the parameters of the model, except those related to the microwave field.
Then, we use those parameters to reproduce the ESR spectra of Fig.~\ref{fig:Fig1}(c). 

Despite the rather complex model involving a number of different interactions with the environment, the amount of parameters 
in the model remains manageable. A summary of the parameters, which we will use in the following, is given in Table~\ref{tab-params}. 
The parameters that are relevant the tunneling current (marked ``Ref.'' in Table \ref{tab-params}) can be extracted from 
reference spectra without microwaves. The other parameters (marked ``Data'' in Table \ref{tab-params}) can be extracted from 
the ESR signal as well as the background signal. This provides enhanced consistency in determining the model parameters for a 
quantitative analysis of the ESR data.

We describe this procedure in more detail in the remainder of this section. 

\begin{table}
    \begin{tabular}{|| c | c | c | r ||} 
    \hline
    {\bf Symbol} & {\bf Parameter} & {\bf Fitted value} & {\bf Source}\\  \hline\hline
    $p$ & tip polarization & $-0.8$ & Ref. \\ \hline
    $\tau_0$ & transmission & $1.85 \times 10^{-5}$ & Ref. \\ \hline
    $\tau_1/\tau_0$ & DOS linear coefficient & $-2.7$\,eV$^{-1}$ & Ref. \\ \hline
    $\tau_2/\tau_0$ & DOS quadratic coefficient & $5$\,eV$^{-2}$ & Ref. \\ \hline
    $\lambda $ & spin-flip parameter & $0.46$ & Ref.  \\ \hline
    $\hbar \Gamma_0$ & intrinsic decay rate & $9$\,neV & Ref. \\ \hline\hline
    $T$ & Temperature & 310\,mK & Input\\ \hline
    $\omega_r$ & Microwave frequency & 61\,GHz & Input \\ \hline\hline
    $\omega_\text{max}$, $B_\text{max}$ & ESR peak position & $V$ dependent& Data\\ \hline
    $V_\text{ac}$ & Microwave amplitude & 19.9\,mV &  Data\\ \hline
    $\hbar\Omega$ & Rabi frequency & 29\,neV & Data\\ \hline
    $\hbar\gamma_0$ & intrinsic coherence rate & 70\,neV & Data\\ \hline
     $\theta$ & homodyne angle & 4.5$^{\circ}$ & Data\\ \hline
    \end{tabular}
    \caption{\textbf{Model parameters} The parameters that are related to the tunneling current can be determined independently 
    from a reference spectrum measured without microwaves (Ref.), the remaining parameters are determined from the ESR spectrum 
    itself (Data). The temperature $T$ and the microwave frequency $\omega_r$ are treated as in input parameters (Input), 
    which are given by the experimental setup. They are normalized such that the parameter $\tau_{0}$ provides the corresponding 
    channel transmission at the Fermi level. The microwave couples differently to the tunneling current and to the spin system. 
    As the relation between these coupling mechanisms is still under debate, we use two parameters $V_\text{ac}$ and $\hbar\Omega$ 
    to describe the interaction with the microwave.}
    \label{tab-params}
\end{table}

\sidecaptionvpos{figure}{c}
\begin{SCfigure*}[][t]
    \centering
    \includegraphics[width=0.66\textwidth]{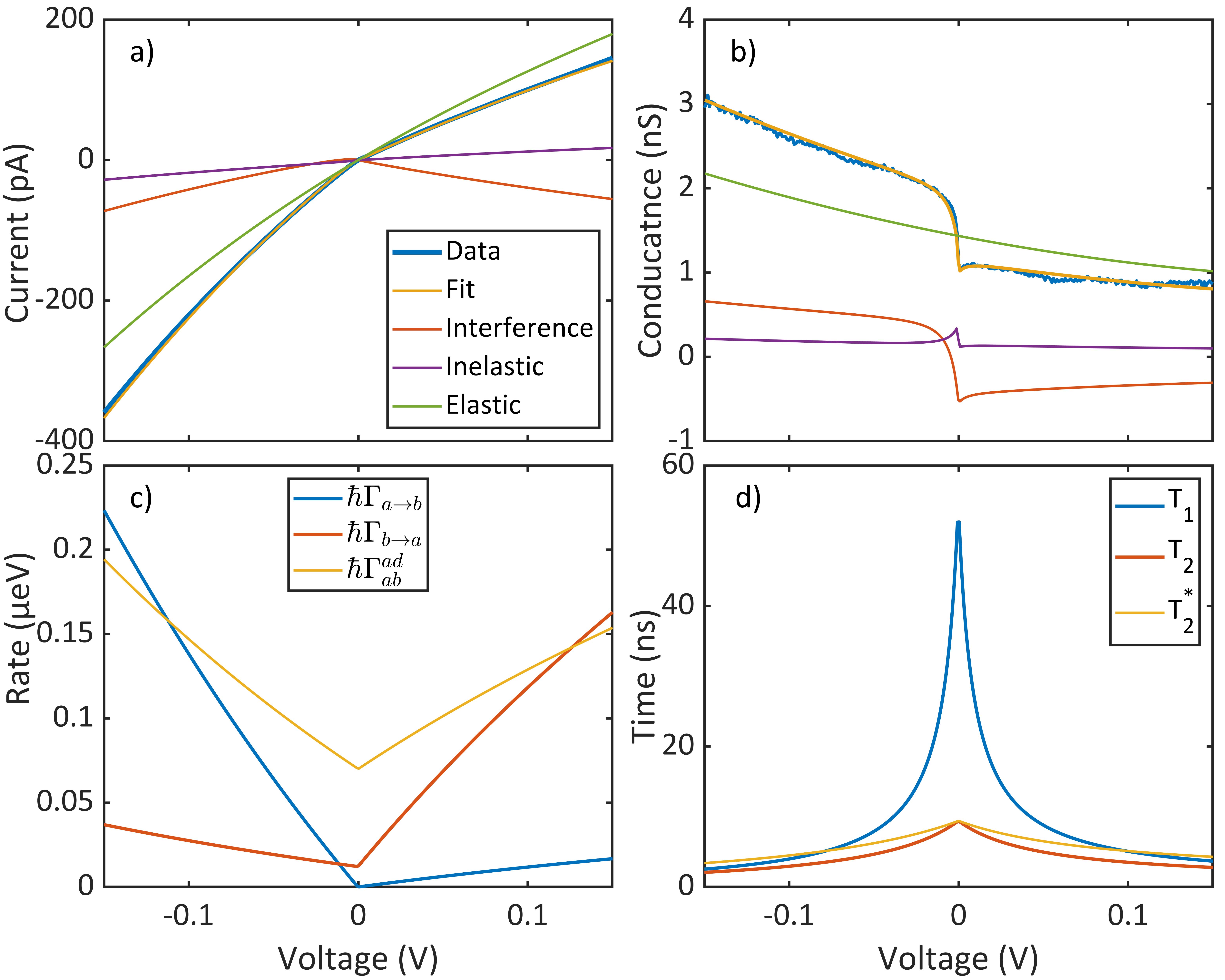}
    \centering
    \caption{\textbf{Reference Measurement without Microwaves:} \panelcaption{a} Current measurement and \panelcaption{b} 
    simultaneous conductance measurement without microwaves as a reference to extract the tunneling parameters. The data is shown 
    in blue and the fit is shown in red. The different contributions to the tunneling current are the elastic term (green), 
    the interference term (yellow) and the inelastic term (purple). \panelcaption{c} Relaxation and decoherence rates for the 
    data set in \panelsubcaption{a}. \panelcaption{d} The corresponding relaxation and decoherence times as defined in 
    Eq.~\eqref{eq-times}.}
    \label{fig:Fig2}
\end{SCfigure*}

\subsection{Current in the absence of microwaves}

In Fig.~\ref{fig:Fig2}(a,b) we show the experimental data for the current and differential conductance, respectively, as a function 
of the dc bias voltage in the absence of microwaves. Notice that the current is nonlinear. It exhibits a conductance step in the 
low bias voltage regime making it asymmetric with respect to the voltage polarity. To model these current-voltage characteristics we shall
take into account two features of the lead Green's functions defined in Eqs.~\eqref{eq-g+-}-\eqref{eq-g-+}. The first feature is an 
energy-dependent DOS and the second feature is a spin-polarized DOS. Assuming that the lead Green's functions are diagonal in spin 
space, they are given by
\begin{eqnarray}
g^{+-}_{j \sigma}(E) & = & 2 \pi i \varrho_{j \sigma}(E) f(E) , \\ 
g^{-+}_{j \sigma}(E) & = & 2\pi i \varrho_{j \sigma}(E) \left[ f(E) - 1 \right] ,
\end{eqnarray} 
where $\varrho_{j \sigma}(E)$ is the DOS of electrode $j=L,R$ for spin $\sigma = \uparrow$,~$\downarrow$ and $f(E) = [1 + 
\exp(E/k_\text{B}T)]^{-1}$ is the Fermi function. Then, to incorporate those two features, we assume that the substrate's DOS is 
constant and spin-independent ($\varrho_{L\sigma}(E) \approx \varrho_L (0)$, where the Fermi level is at $E=0$), while the DOS 
of the right electrode (STM tip) is given by
\begin{equation} \label{eq-DOSR}
    \varrho_{R \sigma}(E) \approx \left( 1\pm p \right) \left[ \varrho_R(0) + \varrho_R^{\prime}(0) E +
    \frac{1}{2} \varrho_R^{\prime \prime}(0) E^2 \right] .
\end{equation}
Here, $p \in [-1,1]$ is the spin polarization of the tip, which is responsible for the current polarization $I_\text{pol}$ being 
nonzero. The $+$ sign applies to $\sigma = \uparrow$, while the $-$ sign to $\sigma = \downarrow$. The different terms in the 
expansion of Eq.~(\ref{eq-DOSR}) can be cast into an energy dependent transmission coefficient $\tau(E) = \tau_0 + \tau_1 E + 
\tau_2 E^2$ with the coefficients $\tau_0 = 4\pi^2 t^2 \varrho_L(0) \varrho_R(0)$, $\tau_1 = 4 \pi^2 t^2 \varrho_L(0) 
\varrho_R^{\prime}(0)$ and $\tau_2 = 2\pi^2 t^2 \varrho_L(0) \varrho_R^{\prime \prime}(0)$. The transmission determines the 
magnitude of the elastic current, see Eq.~(\ref{eq-Iel}). Note that $\tau_0$ is just the transmission coefficient at the Fermi 
level that determines the constant conductance. We also point out that the symmetry of the problem in principle allows us to exchange the densities of states in tip and sample without changing the result as long as the spin polarization is energy independent. Therefore, to lowest order, the actual energy dependence of the individual densities of states is not relevant, as long as the convolution matches the data.

Using these assumptions for the electronic structure of the electrodes, the temperature of the experiment ($T = 0.31$ K), and the 
current formula in the absence of microwaves described in Sec.~\ref{sec-TLS-current} in Eqs.~\eqref{eq-Idark}-\eqref{eq-Iinel}, we 
are able to quantitatively reproduce the experimental results, see Fig.~\ref{fig:Fig2}(a,b). The different current contributions 
(elastic, interference, and inelastic) are shown in Fig.~\ref{fig:Fig2} as well. As expected, the elastic term (green) gives the 
largest contribution followed by the interference term (yellow) and then the inelastic term (purple). Although visible in both panels, 
this is best seen in the conductance spectrum (Fig.~\ref{fig:Fig2}\panel{b}). The elastic term evolves smoothly retracing the 
non-constant DOS. The most prominent feature in the differential conductance is a big step near zero bias voltage, which originates 
from the interference term. This is due to the rather strong spin polarization in the tip, which transfers its polarization onto 
the spin system (see discussion in Appendix \ref{sec-app1}). It is worth stressing that we find a very high spin-polarization 
($|p|=0.8$), which may explain the difficulty in preparing ESR-active tips. Let us also say that, although not featureless, the 
inelastic term only gives a small contribution to the overall signal. From this fit we obtain most of the parameters of the model 
except those related to the microwaves. The values of those parameters are summarized in Table~\ref{tab-params}. 

\begin{table}[b]
    \begin{tabular}{|| c | l ||} 
    \hline
    {\bf Parameter} & {\bf Impact} \\  \hline\hline
    $p$ & $dI/dV$ (inelastic step height) \\ \hline
    $\tau_0$ & $dI/dV$ (overall scaling) \\ \hline
    $\tau_{1,2}/\tau_0$ & $dI/dV$ (overall slope) \\ \hline
    $\lambda $ & $dI/dV$ (inelastic step height vs.\ overall scaling) \\ \hline
    $\hbar \Gamma_0$ & $dI/dV$ (step width) \\ \hline
    $\omega_\text{max}$, $B_\text{max}$ & Map (ESR peak position)\\ \hline
    $V_\text{ac}$ & Background (scaling)\\ \hline
    $\hbar\Omega$ & ESR peak height (overall scaling)\\ \hline
    $\hbar\gamma_0$ & ESR peak width (zero bias voltage offset)\\ \hline
    $\theta$ & ESR spectrum near zero bias votlage\\ \hline
    \end{tabular}
    \caption{\textbf{Impact of Model Parameters} Summary of the parameters used in the modeling and their impact on different 
    measured quantities. Every parameter has a rather unique impact on one feature of the different measured quantities. This 
    allows for a rather consistent fit, which little interdependence between fit parameters.}
    \label{tab-impact}
\end{table}
\sidecaptionvpos{figure}{c}
\begin{SCfigure*}
    \centering
    \includegraphics[width=0.66\textwidth]{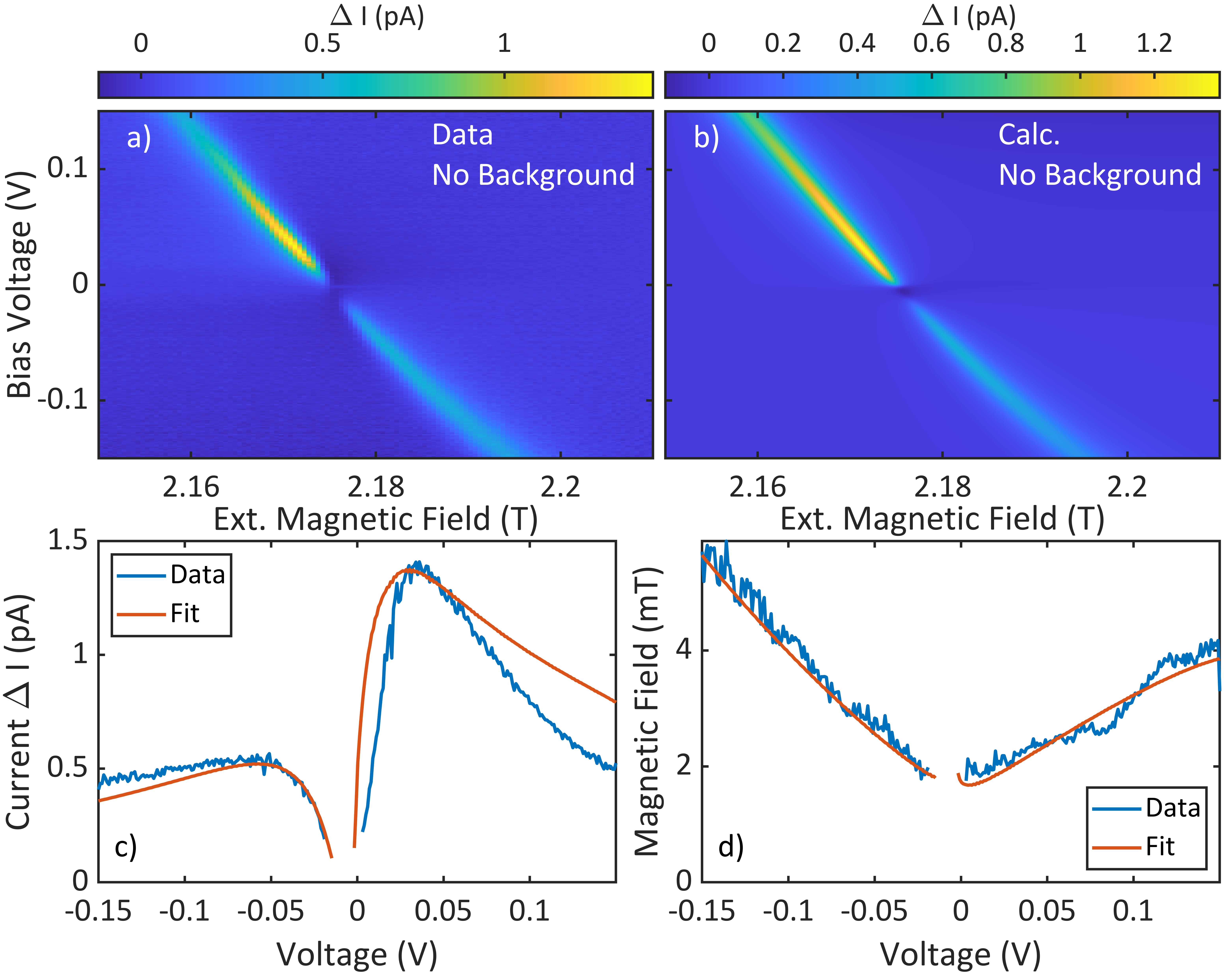}
    \centering
    \caption{\textbf{Analyzing the ESR peak shape:} \panelcaption{a} Measured ESR signal as function of bias voltage and external 
    magnetic field at a constant frequency (61\,GHz) with the off resonance background subtracted from Fig.\ \ref{fig:Fig1}\panelsubcaption{c}.
     \panelsubcaption{b} Calculated ESR signal with the off resonance background subtracted. The parameters are chosen for a best fit to 
     the data in \panelsubcaption{a}. Comparison of the ESR peak height \panelcaption{c} and ESR peak width \panelsubcaption{d} extracted 
     from the data in \panelsubcaption{a} and \panelsubcaption{b}. The fit parameters have been determined from panels \panelsubcaption{c} 
     and \panelsubcaption{d}.}
    \label{fig:Fig3}
\end{SCfigure*}

\subsection{ESR spectra}

The ESR calculation in Fig.~\ref{fig:Fig1}\panel{d} has been done using the parameters extracted from the data without microwaves 
in Fig.~\ref{fig:Fig2}. The relaxation and decoherence rates determining the spin dynamics can be readily calculated from the 
extracted parameters in Fig.~\ref{fig:Fig2}\panel{a} and are displayed in Fig.~\ref{fig:Fig2}\panel{c}. The corresponding time scales 
are plotted in Fig.~\ref{fig:Fig2}(d). For this calculation, we have to include the microwave frequency as an input parameter, which 
we set to 61\,GHz here. The remaining parameters can be extracted from measurements with microwaves. The peak positions $\omega_\text{max}$, 
$B_\text{max}$ are extracted as function of bias voltage. The functional dependence of the peak position on the bias voltage is due 
to the electric field in the tunnel junction induced by the bias voltage, which is discussed elsewhere and will not be detailed here
\cite{kot_electric_2022}. The peak positions are used as input parameters in the calculations. 

For the subsequent analysis, we subtract the off-resonance background in the measured and calculated data in 
Fig.~\ref{fig:Fig1}\panel{c} and \panel{d}, respectively. We take a slice as function of bias voltage at a constant magnetic field that 
does not show any resonance signal and subtract it from every other slice in the data set. The corrected measured and calculated data 
is shown in Fig.~\ref{fig:Fig3}\panel{a} and \panel{b}, respectively, with a smooth background demonstrating that the subtracted 
off-resonance background is independent of magnetic field. With the ESR peak positions known, we can extract the peak height and 
the peak width (full width at half maxiumum) from the corrected data in Fig.~\ref{fig:Fig3}\panel{a}, which are shown in 
Fig.~\ref{fig:Fig3}\panel{c} and \panel{d}, respectively. The data is shown in blue, while the fit is shown in red. The only free 
parameters in this fit are the intrinsic coherence rate $\gamma_0$, which is mostly given by the coupling to the substrate, as well as 
the microwave amplitude $\Omega$ (Rabi frequency), that excites the spin system. Their values are given in Table \ref{tab-params}. 
We find that all parameters have more or less impact on all different quantities, but some quantities are particularly influenced by 
certain parameters, which allows us to determine their values rather precisely. The impact is summarized in Table \ref{tab-impact}. 

\sidecaptionvpos{figure}{c}
\begin{SCfigure*}
    \centering
    \includegraphics[width=0.66\textwidth]{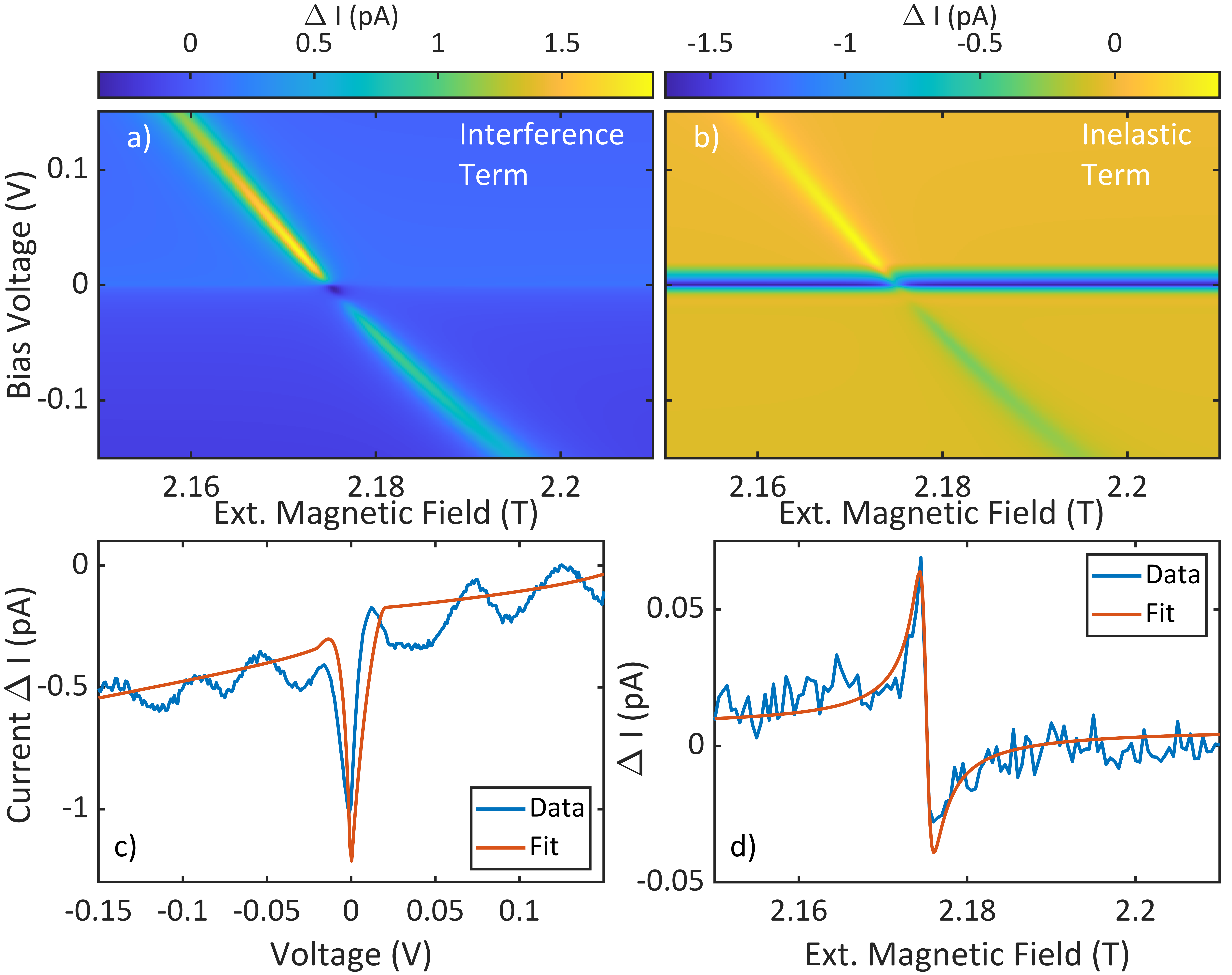}
    \caption{\textbf{Components of the ESR signal:} Contribution of the interference term \panelcaption{a} and the inelastic term 
    \panelcaption{b} to the ESR signal in Fig.~\ref{fig:Fig1}\panelsubcaption{d}. Most of the ESR signal is carried by the interference 
    term. Note that there is a small contribution from the inelastic term to the ESR signal. \panelcaption{c} Off resonance background 
    taken from Fig.~\ref{fig:Fig1}\panelsubcaption{c} and \panelsubcaption{d} for the experimental and the calculated data, respectively. 
    The wavy modulation in the data is due to the microwave interaction with small modulations in the DOS of the leads.
     \panelcaption{d} The homodyne signal near zero bias voltage at $V_\text{bias} = -2.713\,$mV. The angle between the tip spin and 
    the spin system is $\theta = 4.5^{\circ}$. The calculated spectrum was offset by 10\,fA to compensate for slight differences in the 
    background signal.}
    \label{fig:Fig4}
\end{SCfigure*}

The dynamics of the system due to the tunneling current is entirely determined by the parameters that were already independently 
determined by the reference spectrum. We find excellent agreement between the data and the model calculations for the peak height 
and the peak width in Fig.~\ref{fig:Fig3}\panel{c} and \panel{d}. The only deviation we find is in the peak height for voltages 
larger than about 60\,mV, which could be resolved by making some parameters voltage dependent. However, for the proof of principle here, 
we conclude that we have good agreement between data and theory. The corresponding relaxation rates $\Gamma_{a \to b}$ 
and $\Gamma_{b \to a}$ as well as the decoherence rate $\Gamma^{\mathrm{ad}}_{ab}$ are displayed in Fig.~\ref{fig:Fig2}\panel{c}. 
Notice that the impact of the tunneling current is minimized at zero bias voltage.

As before, we separate the contributions to the ESR signal into the three tunneling channels. The elastic current channel does not interact 
with the spin system and, will therefore not contribute to the ESR signal. However, the elastic current channel is influenced by 
the microwave, so that it will contribute to the off-resonance background. The interference term carries the majority of the ESR 
signal as can be seen in Fig.~\ref{fig:Fig4}\panel{a}. The background in the interference term is relatively smooth apart from a 
small step at zero bias voltage. The inelastic term contributes mostly to the off-resonance background with a sizeable feature at 
zero bias voltage as shown in Fig.~\ref{fig:Fig4}\panel{b}. However, the inelastic term also carries part of the ESR signal. This 
ESR signal is substantially smaller than in the interference term, but it cannot be neglected in the data analysis if quantitative 
agreement is to be achieved. 

Another important parameter to be determined is the microwave amplitude $V_\text{ac}$ for the tunneling current. This can best be 
extracted from the off-resonance background, which is shown in Fig.~\ref{fig:Fig4}\panel{c}. The nontrivial behavior of the 
off-resonance background is an indication of a non-constant DOS. A constant/linear/quadratic DOS results 
in a zero/constant/linear offset in the off-resonance background. The sharp dip at zero bias voltage is due to the step feature in 
the interference term (cf.\ Fig.~\ref{fig:Fig2}\panel{b}). The smaller wavy features are higher order details in the DOS, 
which are not captured by the quadratic approximation in Eq.~\eqref{eq-DOSR}. The behavior of the off-resonance background
also depends sensitively on the microwave amplitude, so that we can extract a value of $V_\text{ac}=19.9\,$mV.  

Near zero bias voltage, a small asymmetric ESR peak remains, which is attributed to homodyne detection
\cite{bae_enhanced_2018,yang_coherent_2019,willke_probing_2018,seifert_longitudinal_2020}. The presence of this small signal, which 
can be seen in Fig.~\ref{fig:Fig4}\panel{d}, indicates a finite angle between the tip spin and the spin system resulting in a 
homodyne signal (cf.\ Sec.~\ref{sec-TLS-homodyne}). To determine this angle, we identify the zero crossing of the ESR peak height 
from the calculated spectra at $V_\text{bias} = -2.713$\,mV. The corresponding ESR signal is shown in Fig.~\ref{fig:Fig4}\panel{d}. 
We then set the angle to  $\theta=4.5^{\circ}$ to fit to the data showing excellent agreement. We thus find an overall consistent 
model that quantitatively explains the ESR features and the background signal over several orders of magnitude in the tunneling current. 

\section{What drives the transitions?} \label{sec-EM}

\begin{figure*}[t]
\includegraphics[width=1\linewidth]{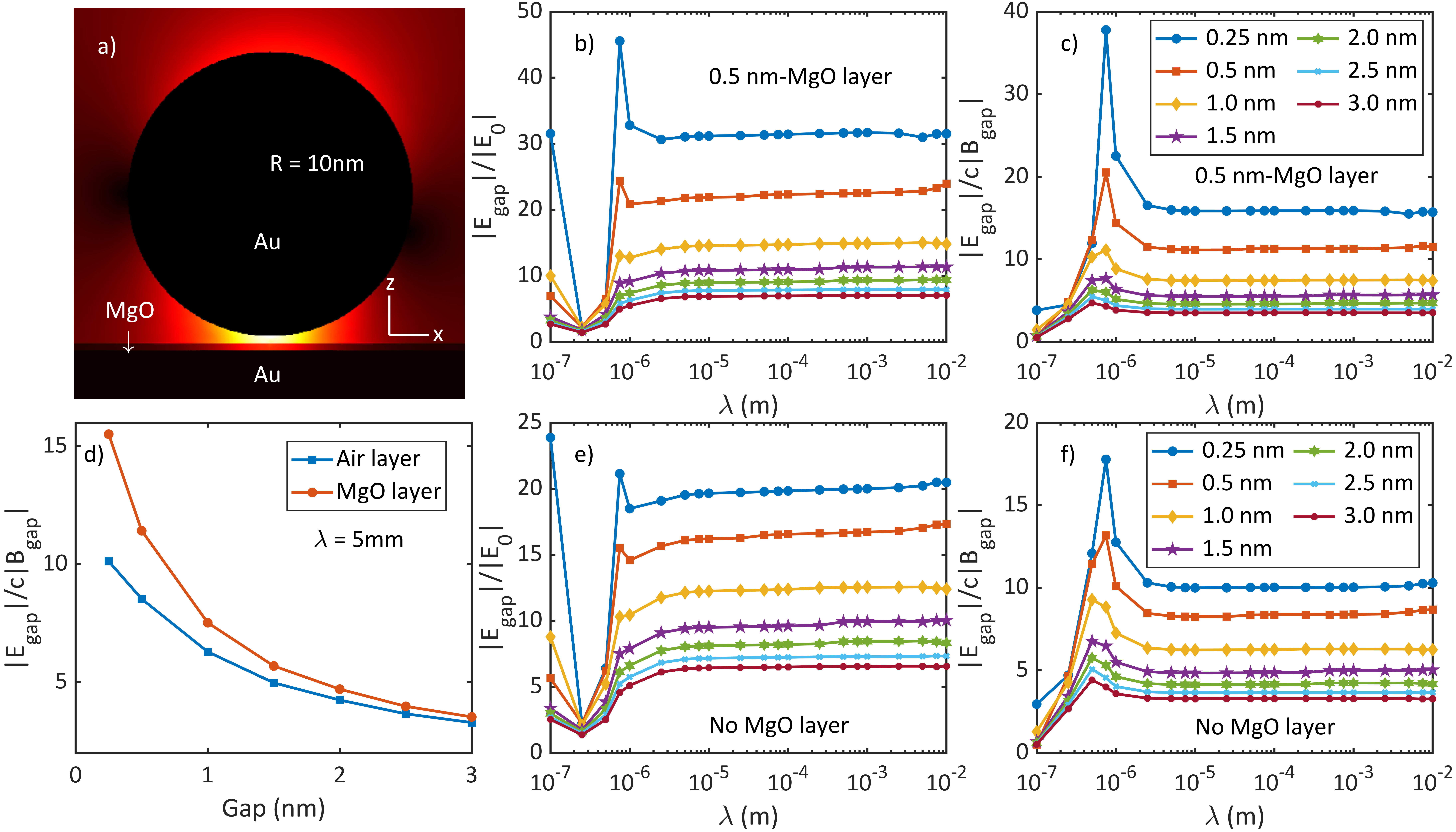}
    \centering
    \caption{\textbf{Electromagnetic simulations:} \panelcaption{a} System considered in the electromagnetic simulations in which a
    metallic sphere is placed above a metallic substrate that features an $0.5$ nm-thick MgO layer on top. The metal is assumed to be 
    gold and the sphere radius is 10 nm. In our simulations we vary the gap size (sphere-MgO distance) between $0.25$ and 3 nm. In this 
    panel we show the spatial distribution of the electric field amplitude in linear scale, from minimum (black) to maximum (yellow), 
    when the system is illuminated with a plane wave with a wavelength of 5 mm, polarization along the $z$-axis and incoming from the 
    negative $x$-direction. \panelcaption{b} Field enhancement ratio $|E_\text{gap}|/|E_0|$ as a function of the radiation wavelength 
    for the system shown in panel \panelcaption{a}. Here, $|E_\text{gap}|$ is evaluated in the middle of the gap region. The different 
    curves correspond to different values of gap size, as indicated in the legend of panel \panelcaption{c}. \panelcaption{c} The 
    corresponding results for the field ratio $|E_\text{gap}|/(c|B_\text{gap}|)$. \panelcaption{d} The field ratio as a function of the 
    gap size for the cases with and without a MgO layer for a wavelength of 5 mm. \panelcaption{e,f} The same as in panels 
    \panelcaption{b,c}, but replacing the MgO layer by air. The gap size indicated in the legend of panel \panelcaption{f} is measured 
    from the top of the $0.5$ nm-thick air layer to the gold sphere to make a fair comparison with the results of panels \panelcaption{b,c}.}
    \label{fig:Fig5}
\end{figure*}

Our theory has been able to clarify many basic questions so far, but we still need to address the most important question in this 
field: what drives the transitions between electronic states? It is somehow surprising that an all-electric ESR can work at all. 
After all, the electric field should not couple to the spin (electric dipole transitions are forbidden due to selection rules). It 
has been argued that the ac magnetic field that accompanies the oscillating electric field is not strong enough to induce the 
transitions between the spin states \cite{Delgado_Lorente_2021}. For this reason, a plethora of alternative mechanisms have been 
proposed \cite{baumann_electron_2015,seifert2020longitudinal,lado2017exchange,ferron2019single,galvez2019cotunneling,
reina_all-electric_2021,reina2023many}, but the community has not yet reached a consensus. In the context of our experiments, in 
which microwaves are supplied by an antenna radiating into the vacuum towards the tunnel junctions, the question boils down to whether 
the ac magnetic field at the junction is sufficiently intense to justify the value of the Rabi frequency $\Omega$ that we have deduced 
from our fits to the experimental results. To shed light on this question, we have carried out classical electromagnetic simulations 
using the frequency-domain, finite-element solver implemented in Comsol Multiphysics. Mimicking recent models for STM 
luminescence \cite{martin2020_unveiling}, we simulate our experiments through an Au nanocavity (with a radius of 10\,nm) separated 
by a gap $d$ from a Au surface, which features a 0.5\,nm thick MgO layer on top, see Fig.~\ref{fig:Fig5}(a). To model Au, we used a 
permittivity obtained from a Drude multi-Lorentz fit to the experimental data of Ref.~\cite{Rakic1998_optical}. Let us stress 
that the choice of the metal here is simply for convenience as all metals behave very similarly in the microwave region. In order to 
determine the electric and magnetic fields in the junction region, we have simulated the propagation of a plane wave incoming along 
the $x$-direction (cf.\ Fig.~\ref{fig:Fig5}(a)) with a wavelength $\lambda$, a far-field magnitude of the electric field equal to 
$|E_0|$, and with a polarization along the axis of the junction, which is the $z$-direction in Fig.~\ref{fig:Fig5}(a). The figure 
shows the electric field distribution for a gap of $d = 0.5$\,nm, a 0.5\,nm thick MgO layer, and an incoming wave of $\lambda = 5$\,mm 
wavelength ($\sim 61$\,GHz). Notice that the electric field is enhanced in the gap region. It is basically constant across the gap 
region and barely penetrates inside the metallic regions. 

To characterize the intensity of the electromagnetic fields in the junction region, we present in Fig.~\ref{fig:Fig5}(b-f) results
for the ratios $|E_\text{gap}|/|E_0|$ and $|E_\text{gap}|/(c|B_\text{gap}|)$, where $|E_\text{gap}|$ and $|B_\text{gap}|$ are the 
magnitudes of the electric and magnetic field, respectively, evaluated half way from the nanocavity and the MgO layer ($c$ is the 
speed of light). Moreover, to understand the role of the MgO substrate, we also present in Fig.~\ref{fig:Fig5}(d,e,f) results without 
the MgO layer, which is simply replaced by an air layer of the same thickness. Let us first discuss the results shown in 
Fig.~\ref{fig:Fig5}(b,e), which shows the electric field enhancement with and without the MgO layer as a function of the radiation 
wavelength ranging from UHV light to the microwave range. The wavelength in our experiments is $\sim 5\,$mm, but to connect with 
field enhancements in other contexts like nanoplasmonics, we present several orders of magnitude in $\lambda$. The different curves 
correspond to different values of the gap size $d$, which are listed in Fig.\ \ref{fig:Fig5}(c,f). Notice that for the smallest gaps 
and microwave wavelengths, the electric field is locally enhanced at the junction region by a factor between 10 and 30 depending on 
the precise value of the gap and the presence or absence of the MgO layer. This enhancement in the microwave region is due to the 
well-known lightning-rod effect and has nothing to do with the excitation of local plasmons, which do not exist in this region of 
the electromagnetic spectrum. Indeed, for all gap sizes, the field enhancement is maximum in the near-infrared and the
red region of the optical window (600-900\,nm). At these wavelengths, the structure supports bright localized plasmons, which couple 
efficiently to the incoming radiation. The local electric field enhancement is progressively reduced as the gap size increases. 
Furthermore, the presence of the MgO layer increases the field enhancement.

In Fig.~\ref{fig:Fig5}(c,f), we show the corresponding results for the ratio $|E_\text{gap}|/(c|B_\text{gap}|)$. This ratio is equal 
to one in the far field and it increases in the gap region up to an order of magnitude for small gaps in the microwave region, meaning 
that the magnetic field is relatively reduced with respect to the electric field. Going back to our relevant wavelength of about 5\,nm, 
we summarize the decrease of the ratio $|E_\text{gap}|/(c|B_\text{gap}|)$ as the gap increases in Fig.~\ref{fig:Fig5}(d) for the cases 
with and without the MgO layer. Notice that the presence of a layer of the MgO dielectrics leads to a more rapid decrease of the field 
ratio with increasing gap size.

Let us now use these simulations to connect with the results of Sec.~\ref{sec-comp}. We note that in our experiments we do not have 
access to the values of the fields in the far-field region, i.e., we do not know $|E_0|$ in our experiments. However, we can use the 
fitted value for $V_\text{ac}$ to get an estimate for the magnitude of the electric field in the junction and then use the results of
Fig.~\ref{fig:Fig5} to obtain the value of the magnetic field that determines the Rabi frequency. We can estimate the magnitude of the 
electric field in the gap region as: $|E_\text{gap}| \approx V_\text{ac} / d$. We have checked that in the microwave regime, this simple 
estimate reproduces very well the rigorous result obtained integrating the profile of the $z$-component of the electric field along 
the axis of the junction. Now, assuming a gap of $d = 2$\,nm (to account for the tiny field penetration in the metallic regions) and 
using $V_\text{ac} = 19.9$\,mV, we obtain $|E_\text{gap}| \sim 10^7\,$V/m. Then, using a typical value of $|E_\text{gap}|/(c|B_\text{gap}|) 
\sim 5$ for the field ratio at the junction, we can estimate that $|B_\text{gap}| \sim 6$\,mT. This value has to be compared with the 
magnetic field deduced from the Rabi frequency: $\Omega = \mu_\text{B} |B_\text{gap}|/\hbar$, where we assume that the $g$-factor is
equal to two for our \spinhalf\ system. Note that the relevant magnetic field component relevant is in the $xy$-plane (perpendicular 
to the static magnetic field), which is precisely the one that is reported in the simulations of Fig.~\ref{fig:Fig5}. Using 
$\hbar \Omega = 29$\,neV, we obtain  $|B_\text{gap}| \sim 0.6$\,mT, which is smaller than the estimate above that was obtained using 
ideal conditions in terms of propagation direction and polarization. Thus, \emph{we conclude from these arguments that in our setup, 
the ESR transitions can be attributed to the ac magnetic field of the microwave radiation.}    

\section{Discussion and Conclusions} \label{sec-conclusions}

The discussion of the previous section applies to situations in which the microwaves are supplied via a nearby antenna radiating 
into vacuum. It would be highly desirable to carry out a similar analysis to clarify what happens in those cases in which an
alternating electric field is fed directly to the tip, like in the original experiment \cite{baumann_electron_2015}. In those cases 
we have no statement about the driving of the transitions and cannot exclude other proposed mechanisms. However, we emphasize here that our theory applies irrespective of how the spin system is driven. Furthermore, a number of theoretical works have treated the ESR-STM theory recently
\cite{berggren_electron_2016,shavit_generalized_2019,galvez2019cotunneling,reina_all-electric_2021,reina2023many}, with 
Refs.~\cite{galvez2019cotunneling,reina_all-electric_2021,reina2023many} being closest in spirit to our work albeit with an excitation 
mechanism employing a time-dependent variation of the tunnel barrier induced by the alternating electric driving field. The most 
notable differences to the previous theories are that we went beyond the small amplitude approximation for the microwave, we allowed 
for a non-constant density of states, and we included the background current to achieve the best possible quantitative agreement with 
the experimental data. 

In summary, we have presented a comprehensive theory of ESR-STM based on a combination of nonequilibrium Green's function techniques 
for the calculation of the current and quantum master equations for the description of the spin dynamics. Our theory naturally includes 
the interplay between the microwave field, the spin dynamics and the tunneling electrons. It accounts not only for the ESR signals but 
also for the background current that contains critical information about the underlying physics. This theory can be applied to any 
spin system and its validity has been established here with a comparison with experimental results for a well-known \spinhalf\ system 
(TiH on MgO). The quantitative agreement between theory and experiment was found over many orders of magnitude of the tunneling current 
and in a consistent manner, such that most model parameters were determined with reference spectra in the absence of microwaves. Such a
quantitative agreement allowed us to unambiguously extract the relevant times scales (relaxation and decoherence times). On more general 
grounds, our theory clarifies what the ESR-STM spectroscopy measures by rigorously justifying some of the heuristic formulas that have 
been customarily employed to analyze ESR signals. This rigorous analysis has allowed us, in particular, to determine the degree of 
spin-polarization of the STM tip in our experiments, which is suprisingly high ($\sim$80\%) and could explain the difficulties in 
fabricating ESR-active tips. In addition, we have corroborated how the asymmetry often observed in ESR spectra is due to homodyne detection, 
i.e., to the coupling of the spin precession and the microwaves when the spins are not aligned. Moreover, with the help of first-principle
electromagnetic simulations we have shown for the scenarios, in which the microwaves are supplied by an antenna, the origin of the spin 
transition can be attributed to the ac magnetic field in the junction region generated by the radiation field. Furthermore, our theory 
is constructed in a modular way such that it can be readily generalized in many different ways: more complex spin systems, inclusion 
of hyperfine interactions, analysis of pump-probe experiments, including other environments (such as dynamical Coulomb blockade), the 
analysis of the interplay between ESR and superconducitivity, etc. In this regard, we believe that the theory presented here will help 
to expand the ESR-STM capabilities in the coming years.

\section{Acknowledgments}

The authors would like to thank Susanne Baumann, Andreas Heinrich, Klaus Kern, Sebastian Loth, Sander Otte, Aparajita Singha, and 
Markus Ternes for fruitful discussions. We are grateful to the European Research Council (ERC) for their financial support. This 
study was funded in part by the ERC Consolidator Grant AbsoluteSpin (Grant No.\ 681164). S.P.\ acknowledges the support from the
Condensed Matter Physics Center (IFIMAC) through an IFIMAC’s Master Grants for 2022-2023. A.I.F.-D.\ acknowledges funding from the 
Spanish Ministry of Science, Innovation and Universities through Grants PID2021-126964OB-I00 and ED2021-130552B-C21, as well as the 
European Union’s Horizon Programme through grant 101070700 (MIRAQLS). J.C.C.\ thanks the Spanish Ministry of Science and Innovation 
(Grant PID2020-114880GB-I00) for financial support and the DFG and SFB 1432 for sponsoring his stay at the University of Konstanz as 
a Mercator Fellow.

\appendix

\section{Experimental methods} \label{sec-Methods}

Experiments were done using a commercial Unisoku USM-1300 STM equipped with high frequency cabling and an antenna. The high frequency 
setup allows for driving ESR signals between 60\,GHz to 100\,GHz \cite{drost_combining_2022}. We cleaned a silver Ag(100) single 
crystal in ultra-high vacuum (UHV) through several cycles of Ar\textsuperscript{+} ion sputtering at 5\,kV and annealing at 820\,K. 
Double layer MgO was grown on the clean silver surface by simultaneous evaporation of Mg onto the sample surface, leaking of
O\textsubscript{2} into the UHV space, and heating of the silver substrate. After the MgO growth, we deposited single Fe and Ti 
atoms onto the surface using electron beam evaporators. Furthermore, the sample was kept below 16\,K during Fe and Ti deposition 
to ensure that the atomic species did not form clusters on the surface. The Ti species naturally hydrate due to the residual 
hydrogen gas found in the UHV space \cite{yang_engineering_2017}. To create ESR sensitive tips we picked up between one and ten 
Fe atoms \cite{baumann_electron_2015}. All measurements were done with a setpoint current of 100\,pA at a setpoint voltage of 100\,mV.

\section{More on the spin$1/2$-system} \label{sec-app1}

% \sidecaptionvpos{figure}{c}
% \begin{SCfigure*}
\begin{figure*}
\includegraphics[width=\textwidth]{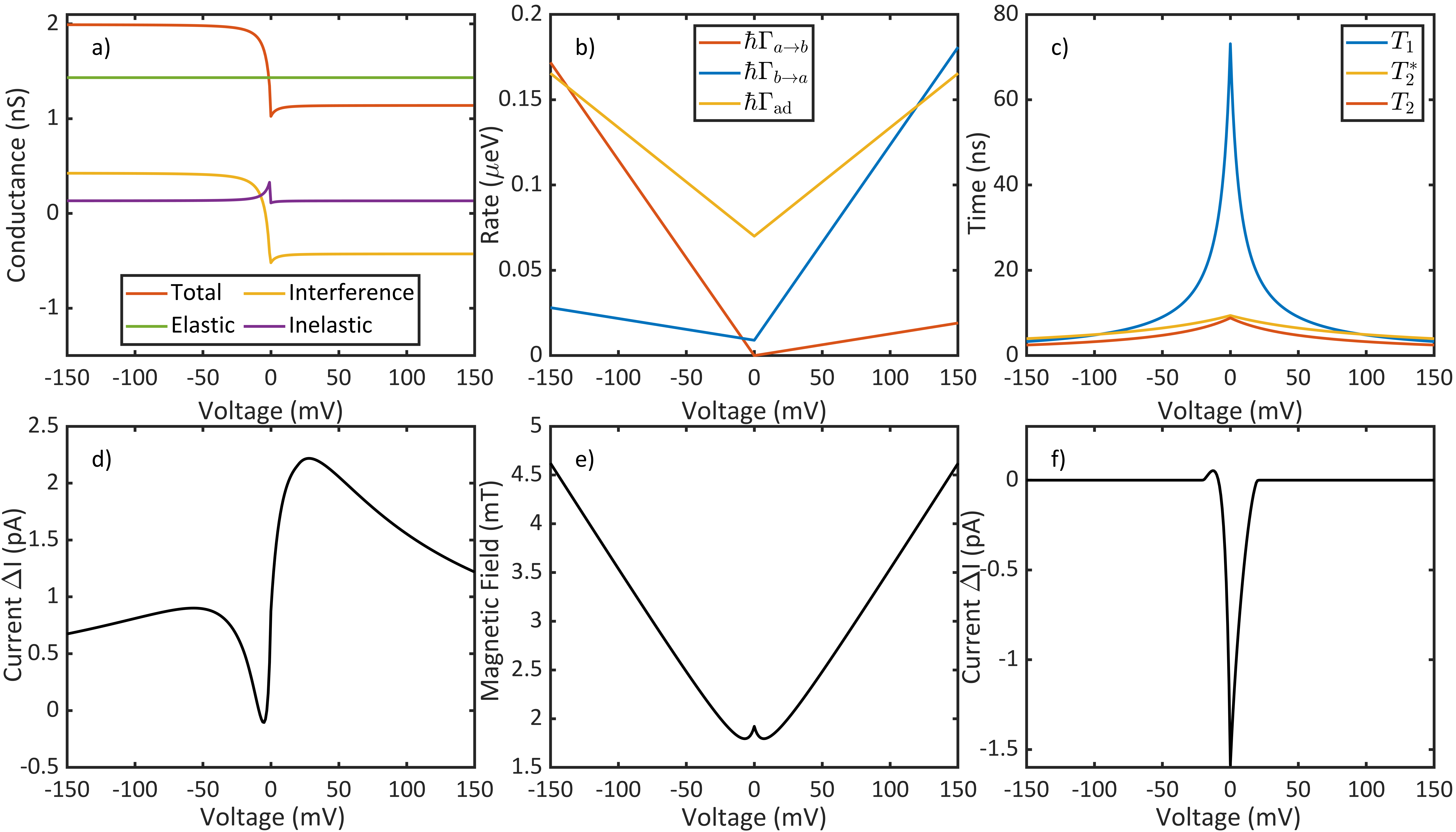}
    \centering
    \caption{\textbf{Analytical results for the \spinhalf\ system:} \panelcaption{a} Conductance in the absence of microwaves. 
    The different contributions are the elastic term (green), the interference term (yellow), the inelastic term (purple), and 
    the total conductance (red). \panelcaption{b} The different rates determining the dynamics of the two-level system. 
    \panelcaption{c} The corresponding relaxation and decoherence times. \panelcaption{d} ESR peak width. \panelcaption{e} ESR 
    peak width. \panelcaption{f} Off-resonant background current. The parameters used to compute these results were those reported 
    in Table~\ref{tab-params}.}
    \label{fig:Fig6}
\end{figure*}
% \end{SCfigure*}

In this Appendix, we elaborate on the case of a \spinhalf\ system and provide some analytical results for more in-depth insight. 
To proceed, we shall assume that the electrodes' DOS are energy-independent with a spin-polarization $p$ and that the bias voltage 
is the dominant energy scale, i.e.\ $\hbar \omega_0 \equiv eV\gg k_B T,~\hbar\omega_{ba}$. Within these approximations the relaxation 
and decoherence rates (and the corresponding characteristic times $T_1$ and $T_2$) are given by
\begin{equation}
    \Gamma_{a \rightarrow b}  =  \frac{\lambda^2 \tau_0}{\pi} \frac{1 + \mathrm{sgn}(\omega_0) p}{2} |\omega_0| ,
\end{equation}
\begin{equation}
    \Gamma_{b \rightarrow a} = \frac{\lambda^2 \tau_0}{\pi} \frac{1 - \mathrm{sgn}(\omega_0) p}{2} |\omega_0| + \Gamma_0 ,
\end{equation}
\begin{equation}
    \Gamma_{ab}^{\mathrm{ad}} = \frac{\lambda^2 \tau_0}{2 \pi} |\omega_0| + \gamma_0 , 
\end{equation}
\begin{equation}
    \Gamma = \frac{1}{T_1} = \frac{\lambda^2 \tau_0}{\pi} |\omega_0| + \Gamma_0 ,
\end{equation}
\begin{equation}
    \gamma = \frac{1}{T_2} = \frac{\lambda^2 \tau_0}{\pi} |\omega_0| + \frac{\Gamma_0}{2} + \gamma_0 ,
\end{equation}
\begin{equation}
    \Gamma^\prime = \Gamma_0 - \frac{\lambda^2 \tau_0}{\pi} p \omega_0 .
\end{equation}
\noindent
With these expressions we can obtain the different elements of the stationary density matrix (cf.\ Eq.~\eqref{eq-rho-str}), which 
adopt the form
\begin{widetext}
\begin{eqnarray}
    \rho_{aa} & = & \frac{\left( |\tilde{\omega}| + \tilde{\gamma}_0 \right) \Omega^2 + 
    \left[ (1 - \mathrm{sgn}(\tilde{\omega}) p) |\tilde{\omega}| + 2 \Gamma_0 \right]  
    \left[ \delta^2 + \left( |\tilde{\omega}| + \tilde{\gamma}_0 \right)^2 \right]}
    {2 \left( |\tilde{\omega}| + \tilde{\gamma}_0 \right) \Omega^2 + 2 \left[ |\tilde{\omega}| + \Gamma_0 \right]  
    \left[ \delta^2 + \left( |\tilde{\omega}| + \tilde{\gamma}_0 \right)^2 \right]} , \\
    \rho_{bb} & = & \frac{\left( |\tilde{\omega}| + \tilde{\gamma}_0 \right) \Omega^2 + 
    2 \left[ (1 + \mathrm{sgn}(\tilde{\omega}) p) |\tilde{\omega}| + 2 \Gamma_0 \right]  
    \left[ \delta^2 + \left( |\tilde{\omega}| + \tilde{\gamma}_0 \right)^2 \right]}
    {2 \left( |\tilde{\omega}| + \tilde{\gamma}_0 \right) \Omega^2 + 2 \left[ |\tilde{\omega}| + \Gamma_0 \right]  
    \left[ \delta^2 + \left( |\tilde{\omega}| + \tilde{\gamma}_0 \right)^2 \right]} , \\
    \tilde \rho_{ab} & = & \frac{\Omega \left( \Gamma_0 - p \tilde{\omega} \right)
    \left[ \delta + i \left(  \tilde{\gamma}_0 +|\tilde{\omega}|  \right) \right]}
    {2 \left( |\tilde{\omega}| + \tilde{\gamma}_0 \right) \Omega^2 + 2 \left[|\tilde{\omega}| + \Gamma_0 \right]  
    \left[ \delta^2 + \left(|\tilde{\omega}| + \tilde{\gamma}_0 \right)^2 \right]} .
\end{eqnarray}
\end{widetext}
Here, we have defined $\tilde{\omega} = \lambda^2 \tau_0\omega_0/ \pi$ and $\tilde{\gamma}_0 = \Gamma_0/2 + \gamma_0$. 
It is interesting to notice that in the limit of large voltages, the stationary density matrix reduces to 
\begin{equation}
    \rho(V \rightarrow \pm \infty) = \frac{1 - \mathrm{sgn}(V)p \sigma^{z}}{2}.
\end{equation}
This means that for sufficiently high bias voltages, the polarization of the spin system follows the spin polarization 
of the tip. In this limit, the expectation value of the magnetization in the $z$-direction becomes $\langle S^z \rangle = 
\mp p \hbar/2$, depending on the sign of the bias. This corroborates the observation that the thermal excitation of the spin system 
for $k_\text{B}T > \hbar \omega_{ba}$, reducing the ESR signal becomes irrelevant at high enough bias voltages making it still 
observable at higher temperatures \cite{seifert_single-atom_2020}.

With these expressions, we can compute the current. First, the elastic current of Eq.~\eqref{eq-Iel} and related quantities 
in the absence of microwaves are given by
\begin{eqnarray}
    I^{\sigma}_\text{el}(V) & = & G_0\tau_0 \frac{1 \pm p}{2} V, \\
    I_{\mathrm{el}}(V) & = & I^{\uparrow}_\text{el}(V) + I^{\downarrow}_\text{el}(V) = G_0 \tau_0 V, \\
    I_\text{pol}(V) & = & I^{\uparrow}_\text{el}(V) - I^{\downarrow}_\text{el}(V) = G_0 \tau_0 p V ,
\end{eqnarray}
where $G_0 = 2e^2/h$ is the conductance quantum. The interference term, see Eq.~\eqref{eq-Iint}, can be written as
\begin{equation}
    I_{\mathrm{int}}(V) = \lambda G_0 \tau_0 p V \left( \rho_{aa} - \rho_{bb} \right),
\end{equation}
while the inelastic current, see Eq.~\eqref{eq-Iinel}, adopts the form
\begin{equation}
    I_{\mathrm{inel}}(V) = \frac{3 \lambda^2 G_0 \tau_0}{4}  V + 
    \frac{\lambda^2 G_0 \tau_0 p}{2}  \tilde{V} \left( \rho_{aa} - \rho_{bb} \right),
\end{equation}
%W
where $\tilde{V} \equiv \sum_{n} J^2_n(\alpha) |V + n \hbar \omega_r /e|$. Thus, the total dark current 
within this approximation is given by
\begin{equation}
    I^\text{dark}(V) = G_0 \tau_0 \left[ \left( 1 + \frac{3\lambda^2}{4} + 
    \frac{\lambda p \Gamma^\prime}{\Gamma} \right)V + \frac{\lambda^2 p \Gamma^\prime}{2 \Gamma}|V| \right] .
\end{equation}
The previous analytical results for the conductance as well as for the relaxation and dechorence rates and time scales are 
illustrated in Fig.~\ref{fig:Fig6}(a-c).

Both the interference and inelastic terms contribute to the ESR lineshape that is given by Eq.~\eqref{eq-ESR} 
where the peak height can be expressed as
\begin{eqnarray}
    I_{\mathrm{peak}}(V) & = & -\lambda G_0 \tau_0 p  \left( V + \frac{\lambda \tilde{V}}{2} \right) 
    \frac{ \Gamma_0 - p \tilde{\omega} }{\Gamma_0 + |\tilde{\omega}|} \times \\
    & & \frac{\Omega^2}{\left( |\tilde{\omega}| + \tilde{\gamma}_0 \right) 
    \left( |\tilde{\omega}| + \Gamma_0 \right) + \Omega^2}, \nonumber
\end{eqnarray}
while the corresponding linewidth adopts the form
\begin{equation}
    {\cal W}(V) = \left( |\tilde{\omega}| + \tilde{\gamma}_0 \right)
    \sqrt{1 + \frac{\Omega^2}{\left( |\tilde{\omega}| + \tilde{\gamma}_0 \right) \left( |\tilde{\omega}| + \Gamma_0 \right) }} .
\end{equation}
These expressions qualitatively reproduce the non-monotonic behavior of the peak height shown in Fig.~\ref{fig:Fig3}(c) and 
the approximate linear dependence of the linewidth with the bias voltage, see Fig.~\ref{fig:Fig3}(d). This is illustrated in 
Fig.~\ref{fig:Fig6}(d,e). However, for a quantitative agreement with our experimental results the energy dependence of the DOS 
cannot be ignored and the full theory descibed in the main text has to be used.

It is worth mentioning that within the approximations used in this appendix the background or off-resonant current is solely due
to the inelastic term and it is given by
\begin{equation}
    I_\text{background}(V) = \frac{\lambda^2 G_0 \tau_0 p}{2} \frac{\Gamma^\prime}{\Gamma} \left( \tilde{V} - |V| \right) .
\end{equation}
This result is illustrated in Fig.~\ref{fig:Fig6}(f).

Finally, it is also interesting to address the homodyne detection discussed in Sec.~\ref{sec-TLS-homodyne}. Assuming a 
constant DOS, Eq.~\eqref{eq-homodyne2} becomes
\begin{eqnarray}
    I_{\mathrm{int}}(V,\alpha) & = &  \lambda G_0 \tau_0 p V \left( \rho_{aa} - \rho_{bb} \right) \cos \theta + \\  
    & & \lambda G_0 \tau_0 p V_\text{ac} \Re\left\{ \tilde \rho_{ba} \right\} \sin \theta \nonumber .
\end{eqnarray}
As discussed in Sec.~\ref{sec-TLS-homodyne}, the second term is responsible for the asymmetry in the ESR spectra.

% \bibliography{New-version/local_bibliography}
% \bibliographystyle{apsrev4-1new}


\end{document}